\documentclass[conference]{IEEEtran}
\IEEEoverridecommandlockouts
\usepackage[ruled,vlined]{algorithm2e}
\usepackage{cite}
\usepackage{amsmath,amssymb,amsfonts}
\usepackage{textcomp}
\usepackage{xcolor}
\usepackage[numbers]{natbib}
\usepackage{pifont}
\usepackage{adjustbox} 
\usepackage{balance}

\usepackage{float}
\usepackage{natbib}
\usepackage{tcolorbox}
\usepackage{array} 
\usepackage{braket}
\usepackage{enumitem}
\setlist[itemize]{topsep= 1pt, partopsep=-0.5pt}
\usepackage[colorlinks=true, citecolor=blue, linkcolor=red]{hyperref}
\usepackage{url}
\usepackage{xurl}

\def\BibTeX{{\rm B\kern-.05em{\sc i\kern-.025em b}\kern-.08em
    T\kern-.1667em\lower.7ex\hbox{E}\kern-.125emX}}

\begin{document}

\title{Space-Time Optimisations \\ for Early Fault-Tolerant Quantum Computation}

\author{\IEEEauthorblockN{Sanaa Sharma}
\IEEEauthorblockA{University of Cambridge} 
Cambridge, United Kingdom \\
ss3241@cam.ac.uk 
\and
\IEEEauthorblockN{Prakash Murali}
\IEEEauthorblockA{University of Cambridge}
Cambridge, United Kingdom\\
pm830@cam.ac.uk}


  
\pagestyle{empty}
\maketitle

\begin{abstract}
Fault-tolerance is the future of quantum computing, ensuring error-corrected quantum computation that can be used for practical applications. Resource requirements for fault-tolerant quantum computing (FTQC) are daunting, and hence, compilation techniques must be designed to ensure resource efficiency. There is a growing need for compilation strategies tailored to the early FTQC regime, which refers to the first generation of fault-tolerant machines operating under stringent resource constraints of fewer physical qubits and limited distillation capacity. Present-day compilation techniques are largely focused on overprovisioning of routing paths and make liberal assumptions regarding the availability of distillation factories. Our work develops compilation techniques that are tailored to the needs of early FTQC systems, including distillation-adaptive qubit layouts and routing techniques. In particular, we show that simple greedy heuristics are extremely effective for this problem, offering significant  reduction in the number of qubits compared to prior works. Our techniques offer results with an average overhead of 1.2$\times$ in execution time for a 53 \% reduction in qubits against the theoretical lower bounds. As the industry develops early FTQC systems with tens to hundreds of logical qubits over the coming years, our work has the potential to be widely useful for optimising program executions.
\end{abstract}

 
\section{Introduction}
\label{introduction}
 
Quantum computers are poised to have practical applications in chemistry, material science, cryptography and other domains~\cite{doi:10.1137/S0097539795293172, PhysRevResearch.3.033055}. There has been tremendous hardware progress, resulting in noisy quantum computers with hundreds of physical qubits~\cite{Acharya2025,Liu2025}. These systems typically have gate (quantum instruction) error rates greater than $10^{-3}$. However, practically useful applications demand qubits with gate error rates below $10^{-10}$. To bridge this gap, quantum error correction (QEC) is required. QEC protects quantum information by encoding the state of a \emph{logical qubit} into several physical qubits and continuously detecting and correcting errors~\cite{PhysRevA.52.R2493}. By combining QEC with \emph{fault-tolerant} operations, we can enable reliable execution of practical quantum applications. 


Early fault-tolerant quantum computers (FTQC) are being developed by physically implementing QEC on real hardware~\cite{Krinner2022,Acharya2025, PhysRevLett.129.030501}. These systems currently have only a handful of logical qubits, with very limited support for running logical operations~\cite{reichardt2025faulttolerantquantumcomputationneutral}. In the next five years, quantum processors are expected to scale to  hundreds of logical qubits, supporting several hundred logical operations, building on current systems that already demonstrate on the order of tens of logical gates \cite{reichardt2025faulttolerantquantumcomputationneutral,google2024roadmap,IBM2024roadmap}. These systems have unique resource constraints such as limited logical qubit counts, restricted logical qubit connectivity and local instruction sets. To support application executions well, we require compilation strategies that are tailored to the capabilities of early FTQC systems.  
 
 To build early and eventually scalable FTQC systems, one of the most important strategies is based on the \emph{surface code} as the underlying choice of QEC. This code represents logical qubits using a square patch or surface of qubits~\cite{Bravyi:1998sy, 10.1063/1.1499754}. 
 It includes a set of physical qubits for storing data and syndrome qubits  for determining errors in the logical qubit by repetitive parity check measurements. Google recently demonstrated surface code-based QEC with one logical qubit and academic groups have demonstrated primitive logical operations~\cite{Acharya2025,  Andersen2020}. However, to realise early scientific applications of quantum computing that are beyond classical computing, we require systems with a few hundred logical qubits~\cite{beverland2022assessingrequirementsscalepractical}. Our work explores how compilation techniques can assist in reducing qubit requirements at this scale.

To reduce qubit requirements, our work focuses on two opportunities. First, logical operations on the surface code are typically local and require very few ancilla qubits ~\cite{Horsman_2012,PhysRevResearch.6.043125}. That is, two-qubit operations can be performed between neighbouring (or diagonal) logical qubit with the help of zero, one or two ancillas depending on the operation type.  
Existing compilation techniques assume that all logical qubits require these ancillas in every timestep, resulting in overprovisioning of ancillas~\cite{Litinski2019gameofsurfacecodes, PRXQuantum.3.020342}. Second, FTQC requires resource-intensive \emph{magic state distillation} factories and associated routing qubits. Distillation factories are required to implement T gates (a type of non-Clifford gate) that cannot be executed directly on the surface code. Magic state qubits produced using distillation must be routed to the logical qubits when required. Existing compilers either ignore distillation and its routing requirements or assume a large number of distillation factories and overprovision routing paths for them~\cite{10.1145/3720416}.

We develop a compiler for early FTQC systems based on greedy heuristics that reduce qubit requirements through two techniques. First, to reduce qubit requirements during logical operations, we provision a limited number of ancilla qubits. Since quantum operation types are known at compile time, we rearrange data qubits and route ancilla qubits to the desired locations. By scheduling these movement operations in advance, we avoid increasing execution time significantly. Second, we insert a limited number of routing paths and use greedy heuristics to route magic states from the distillation factories to the logical qubits where they are used. Since each magic state requires several logical cycles to produce, we use this window to pack as many qubit movement operations as possible to hide movement latency. 
Both techniques rely on determining a minimum set of logical qubit movements either to rearrange ancillas or move magic states. We show that a simple set of greedy heuristics based on a weighted version of Dijkstra's algorithm~\cite{Dijkstra1959} is effective for this problem. 


We develop our compiler with the IBM Qiskit framework and perform  evaluations on a set of early FTQC benchmarks including models for condensed matter simulation and arithmetic circuits. Overall, our contributions are as follows:

\begin{itemize}
\item Across benchmarks, our compiler offers an average 53\% reduction in logical qubit count, with an average execution time increase of 1.2$\times$, compared to layouts given in \cite{Litinski2019gameofsurfacecodes}. In comparison with \cite{10.1145/3720416}, our approach achieves, on average, a 2$\times$ reduction in the spacetime volume with a single distillation factory. Relative to \cite{10946814}, it achieves a 30\% reduction in spacetime volume for selected circuits, with more pronounced improvements at higher factory counts.

\item We show that it is important to jointly optimise the allocation of routing paths in the compiler and the allocation of magic state factories in the architecture. Compared to Littinski's fast block layout \cite{Litinski2019gameofsurfacecodes}, which assumes a large number of distillation factories, our compiler is \emph{distillation-adaptive}. It reduces spacetime volume by allocating just enough routing paths to prevent bottlenecks from magic state routing.
\item Our work enables a richer set of space vs. time tradeoffs compared to prior work which handpicks certain space-time configurations ~\cite{Litinski2019gameofsurfacecodes, 10.1145/3720416, Lao_2019,  PRXQuantum.3.020342}. This is an important capability as we build early FTQC systems, allowing hardware designers to choose configurations that can run on a small number of qubits.

\end{itemize}

\section{Background}

\begin{figure} 
  \centering
  \includegraphics[width=0.5\textwidth]{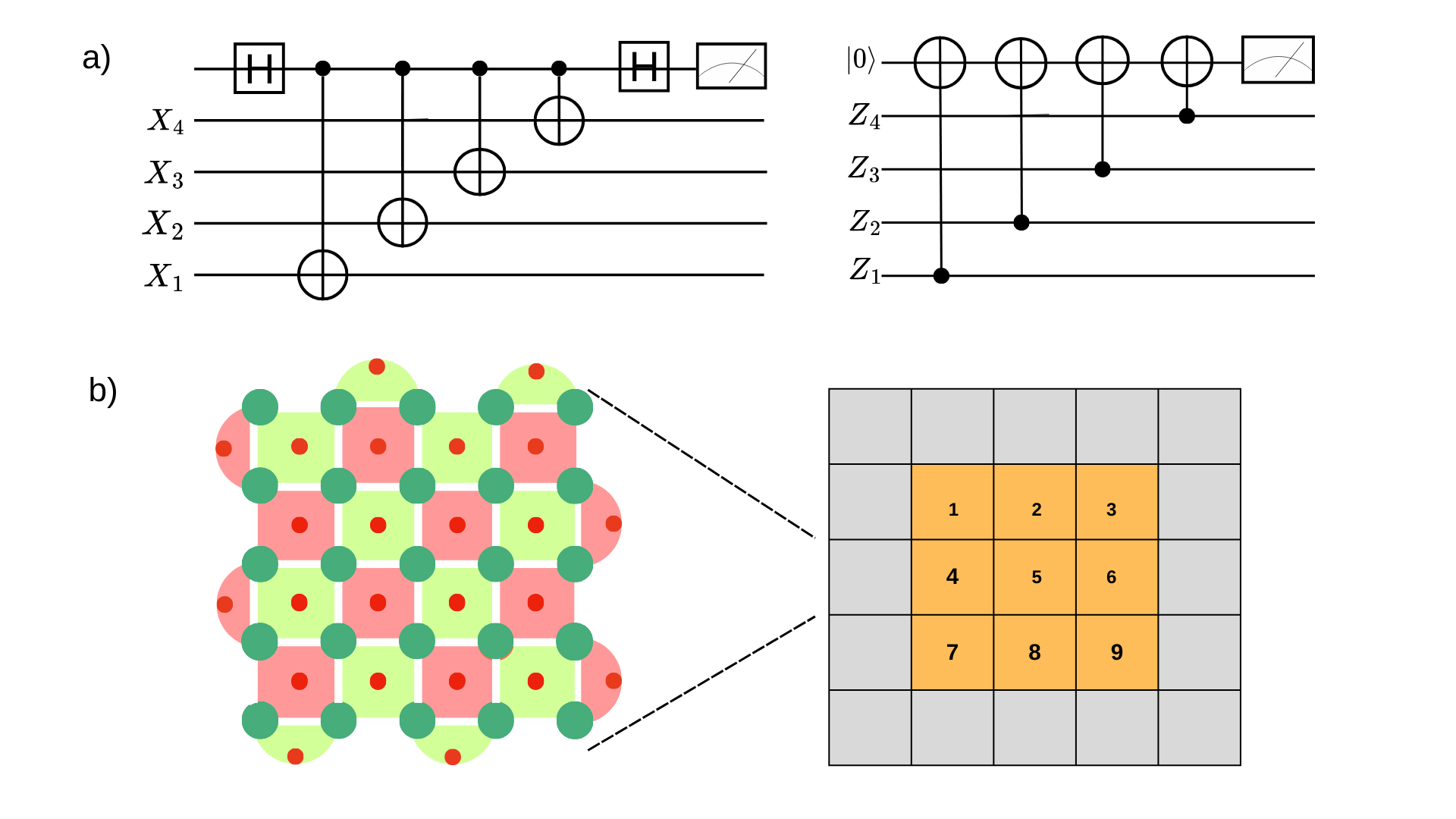}
  \caption{\textbf{(a)} Circuit representation of X and Z syndrome measurements. \textbf{(b)} (Left) The logical qubit has $2d^{2}-1$ physical qubits where d is the code distance (here d=5). The big green circles are the data qubits while the small red circles are the syndrome qubits (red and green faces correspond to the Z and X syndromes, respectively). (Right) Layout of logical qubits in the FTQC system. Orange qubits are the logical data qubits, while the grey region consists of logical ancillas, which are used both for instruction implementation and routing magic states.} 
  \label{fig:2}
\end{figure}

\subsection{Quantum Error Correction and Surface Codes}

Quantum error correction is a technique to suppress errors in quantum computation by encoding a logical qubit with several physical qubits. When QEC is used, program qubits are mapped to logical qubits in the architecture. Each logical qubits is composed of a number of noisy physical qubits. Works such as  \cite{Acharya2025,Bluvstein2024} describe the successful implementations of QEC and small circuit demonstrations on logical qubits.

One approach to building such quantum computers is based on surface codes. These codes only use nearest-neighbour (NN) interactions in physical qubits for error correction and hence can be implemented on a 2D grid of physical qubits. In a logical qubit encoded with surface codes (Figure  \ref{fig:2}(b)), there are two types of qubits, data and syndrome, which differ in their function within the code. The data qubits are used for computation, whereas the syndrome qubits are frequently interacted with to conduct parity measurements. Figure \ref{fig:2}(a) gives the circuit representation of X and Z syndrome measurements, which ensures that the data qubits are in a simultaneous eigenstate of the operators. The errors in the data qubits are detected by the syndrome qubits during computation. A single syndrome measurement constitutes one code cycle. 

Code distance, in error-correcting codes, is the minimum number of physical qubit errors that can cause a logical error. It is also the number of qubits on one side of the logical qubit. In Figure  \ref{fig:2}(a), the code distance $d$ is 5. The larger the code distance, the more errors can be tolerated on the logical level. Code distance affects the execution time of a logical quantum instruction. Each instruction typically requires either $d$ timesteps or some multiple of $d$ to ensure fault-tolerance.

\subsection{Lattice Surgery}
\begin{figure} 
  \centering
  \includegraphics[width=0.5\textwidth]{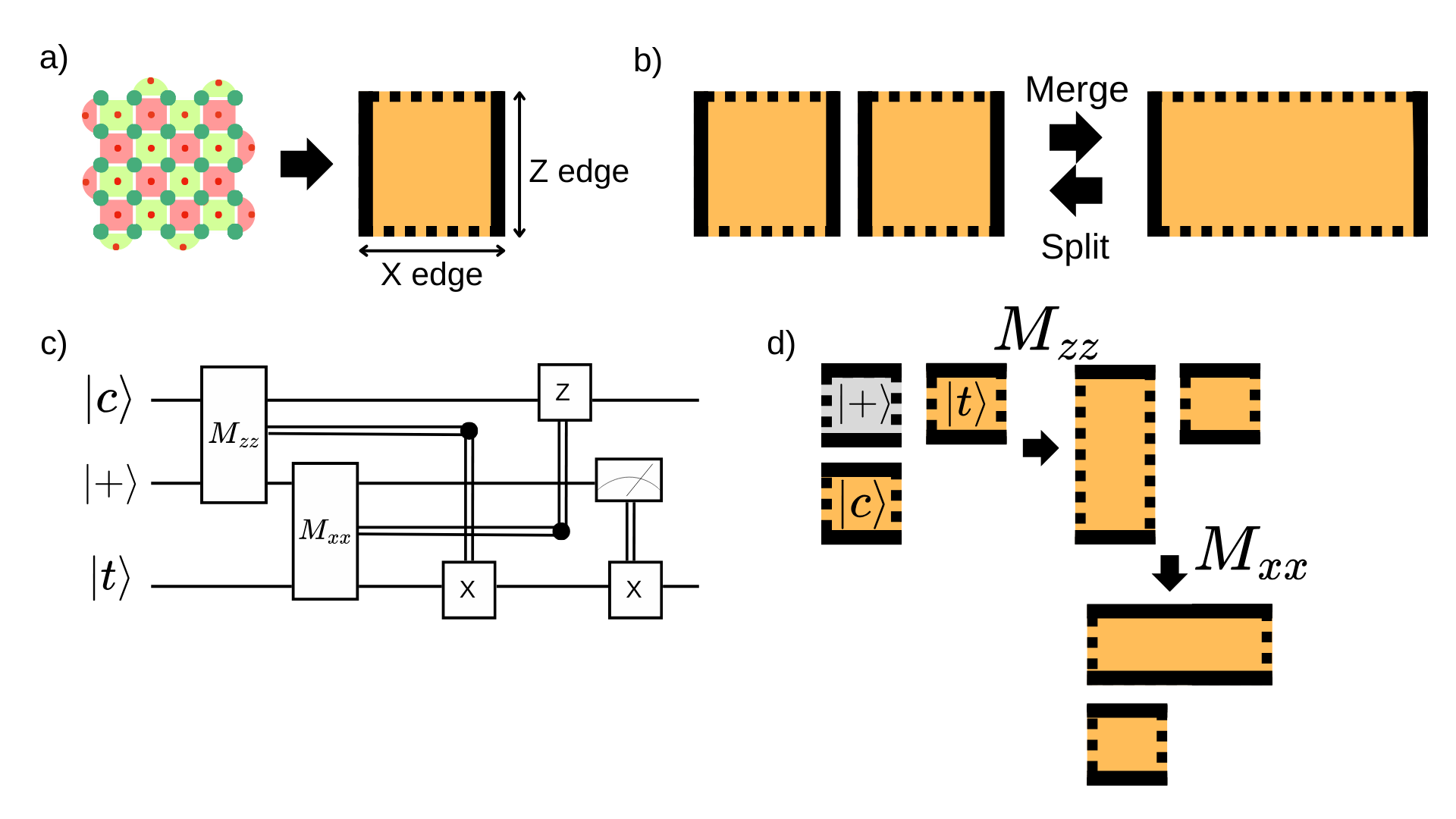}
  \caption{\textbf{(a)} Logical qubit with X and Z edges. \textbf{(b)} Lattice surgery merge and split operations on two logical qubits. \textbf{(c)} CNOT gate decomposed into lattice surgery primitives. It has a control and a target qubit, given by $\ket{c}$ and $\ket{t}$  qubits. The ancilla qubit is $\ket{+}$.\textbf{(d)} CNOT operation implementation on two logical qubits and an ancilla.}   
  \label{fig:5}  
\end{figure}

 Lattice surgery is the standard approach for implementing operations on the surface code \cite{Horsman_2012}. Lattice surgery implements operations non-transversally using the ``merge" and ``split" techniques while maintaining fault-tolerance across the 2D grid. Figure \ref{fig:5}(a) represents a logical qubit as a single qubit cell, with Z and X edges. These edges represent the syndromes on the qubit's boundaries. The merge and split operations, as shown in Figure \ref{fig:5}(b), are the primitive operations that make up the lattice-surgery fault-tolerant operations. Merging is done by measuring joint syndromes across the boundaries of the two qubits during error correction to make a single qubit. Splitting, on the other hand, breaks down a large surface into smaller ones by introducing boundaries. 
 
 Figure \ref{fig:5}(c) is a circuit decomposition of the lattice surgery-based Controlled-NOT (CNOT) operation. CNOT is a common operation in quantum algorithms. The circuit consists of two merge and split operations using measurement operations ($M_{zz} $ and $M_{xx}$), the outcomes of which are used to apply conditional X and Z operations on the control and target qubits. In Figure  \ref{fig:5}(d), the lattice surgery operations along with the placement of control, ancilla and target qubits are described. 
 The execution time in lattice surgery operations refers to the number of timesteps required to execute an operation and is proportional to the code distance of the surface codes employed. Our instruction set is described in Section \ref{sec:evaluation}. 

 \subsection{Magic State Distillation}

The surface code natively supports Clifford gates, a gate set that maps Pauli operators to other Pauli operators under conjugation. Non-Clifford gates 
(\textit{T} gates) are implemented through magic state distillation, making them resource-intensive. To implement a T gate,  a magic state of the form   $\frac{1}{\sqrt{2}}(\ket{0}+e^{i\pi/4}\ket{1})$ is consumed. The natural generation of such states is error-prone, and therefore, a method called \emph{distillation } is used \cite{PhysRevA.71.022316, Litinski2019magicstate}. This method is used to generate higher-fidelity magic states from low-fidelity states via a lengthy protocol conducted separately. 
Once a magic state is prepared outside the computation block, it is routed close to the data qubits, on which the T gate needs to be performed. A standard scheme for  distillation is called 15-to-1 distillation~\cite{PhysRevA.71.022316}.

\section{Related Work}

Given that optimising surface code circuits is known to be NP-hard \cite{Herr2017}, a variety of optimisation techniques at different levels of abstraction have emerged in recent years \cite{Litinski2019gameofsurfacecodes,10.1145/3720416,10.1109/MICRO.2018.00072,10.1145/3466752.3480072, Watkins2024highperformance}. For example, in \cite{Litinski2019gameofsurfacecodes}, the entire quantum circuit is decomposed into a set of Pauli-product rotations and measurements. The method uses commutation rules between any two Pauli gates to reduce a quantum program down to a set of multi-qubit non-Clifford operations and multi-qubit measurements. This method offers a promising solution for circuits with a large set of operations, as it manages to ultimately decompose them into a series of Pauli-product rotations. However, the latency and qubit cost of implementing multi-qubit Pauli product operations is dependent on the underlying hardware and instruction set assumptions and is typically not evaluated~\cite{Moflic:2024xhe}. In other works such as \cite{10.1145/3720416, PRXQuantum.3.020342}, methods are described to optimise paths implementing two-qubit operations and for routing magic states. These methods help define the depth of a particular circuit for certain layouts, which is crucial for performing 2-qubit gates. However, certain bottlenecks, such as distillation processing time, are not accounted for. Other NISQ compilers, such as \cite{10.1145/3470496.3527394, 10.1145/3297858.3304023}, describe techniques for reducing computational overhead but do not address challenges related to distillation routing.

  \section{Problem Statement}
  \label{problem statement}
As mentioned in section \ref{introduction}, surface code compilation faces two major sources of overhead. First, although logical operations on the surface code are inherently local and often require few ancilla qubits, existing compilers liberally assume that ancillas are needed for every logical qubit in every timestep, leading to systematic overprovisioning.   For example, in the compact block of \cite{10.1145/3720416}, the ratio of data to ancilla qubits is 1:2, thereby imposing significant demands on qubit resources. Second, magic state distillation and routing constitute a major bottleneck: distillation factories are highly resource-intensive but essential for implementing non-Clifford gates such as the T gate, and the distilled states must be routed efficiently to their target logical qubits. Existing compilers often either neglect these requirements entirely or operate under the assumption of abundant distillation capacity and routing paths, an assumption that is unrealistic in early fault-tolerant settings. In \cite{10.1145/3720416,  Litinski2019gameofsurfacecodes, 10.1145/3466752.3480072}, distillation bottlenecks are not considered, and the proposed compilation schemes are developed without explicit optimisation for resource-constrained architectures. 

\noindent This motivates the following questions: 
 
\begin{itemize}
     \item Given the critical bottlenecks in surface code compilation, what represents the most resource-efficient balance, and how can optimisation techniques be structured to manage the trade-off between qubit overhead and execution time? 
     \item At the architectural level, how can we incorporate properties of early fault-tolerant systems such as limited distillation capabilities and routing paths during compilation?
 \end{itemize}

 These questions motivate the development of a compiler capable of  optimising trade-offs between qubits and execution time while incorporating architectural constraints that reflect the resource limitations.

\section{Methods}

 \begin{figure}  
  \centering
  \includegraphics[width=0.5\textwidth]{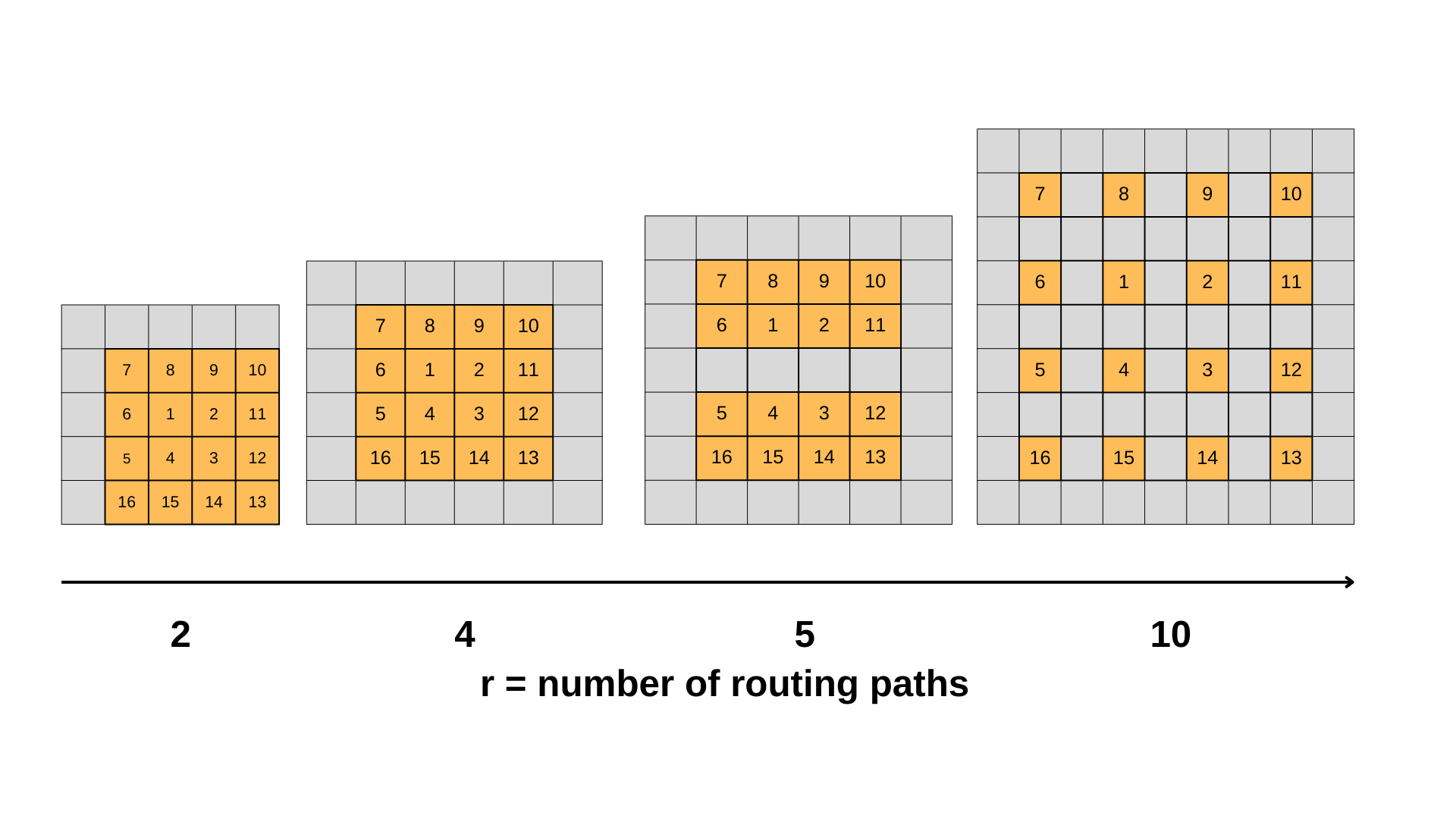}
  \caption{Qubit layouts generated as a function of routing paths. The data qubits are in orange, while the bus qubits are in grey. } 
  \label{fig:9}
\end{figure}

 Our compilation scheme takes as input a quantum program expressed in the Clifford+T gate set. The pipeline comprises three stages: mapping, routing, and scheduling. In the mapping stage, the circuit qubits are assigned to a 2D grid of logical qubits. This grid accommodates both the qubits required for the circuit itself and the additional ancilla qubits necessary to support logical operations and routing. Within the mapping stage, the grid is parameterised by the number of routing paths. These routing paths are created by assigning rows/columns of bus qubits, which can also be used as operational ancilla. In Figure  \ref{fig:9}, a circuit with 16 data qubits (in orange) is shown. The routing paths, depicted in grey, represent the bus qubits placed around the data qubits, going from 2 to 10. For instance, the first subfigure in Figure~\ref{fig:9} includes two edges of bus qubits (top and left), corresponding to two routing paths. The subsequent layout extends this to four routing paths, formed by the four edges surrounding the data qubits.  In comparison to prior works discussed in Section~\ref{problem statement}, where the ratio of data to ancilla qubits is typically 1:2 or 1:3, our layouts (with $r=3,4$) achieve a more resource-efficient ratio of approximately 2:1. We increase routing paths by inserting columns or rows of bus qubits between the data qubits. These paths are required to perform lattice surgery operations, which need special configurations to execute. More routing paths increase the total number of qubits used in compilation but make it easier for operations to run in parallel, consequently reducing the overall time.

  We assign an initial static mapping to our grid depending on the programs. For 2D lattice models, we provide a limited optimisation  at this stage as the total execution time benefits from adopting a layout that is aligned with the gate dependencies in the application. For the remaining benchmarks, we assign qubits based on the number of two-qubit gates shared between qubit pairs. We consider a range of 2D circuits derived from condensed matter models such as the Ising, Heisenberg, and Fermi–Hubbard Hamiltonians. These circuits are important for early FTQC since they represent applications which are scientifically interesting and beyond classical computing. Additionally, their Hamiltonians are defined by nearest-neighbour (NN) interactions between spins, a structure that aligns well with the connectivity constraints of fault-tolerant quantum processors. Logical qubits are laid out on a 2D grid, and entangling operations are typically limited to interactions between adjacent patches (NN interactions). When translated into quantum circuits, these NN interactions are represented by CNOT gates between pairs of qubits, accompanied by additional single-qubit operations.  

 After the initial mapping, primary optimisation is performed during the routing phase, where we employ an adapted version of Dijkstra’s algorithm for efficient path-finding \cite{Dijkstra1959}, combined with gate-dependent moves for neighbouring qubits. For our ancilla-optimised layouts (with $r=3,4$), it is crucial to minimise the number of move operations across the 2D grid during circuit implementation. Our approach ensures minimal disturbance to the 2D layout while maintaining operational efficiency.

 \begin{figure}   
  \centering
  \includegraphics[width=0.5\textwidth]{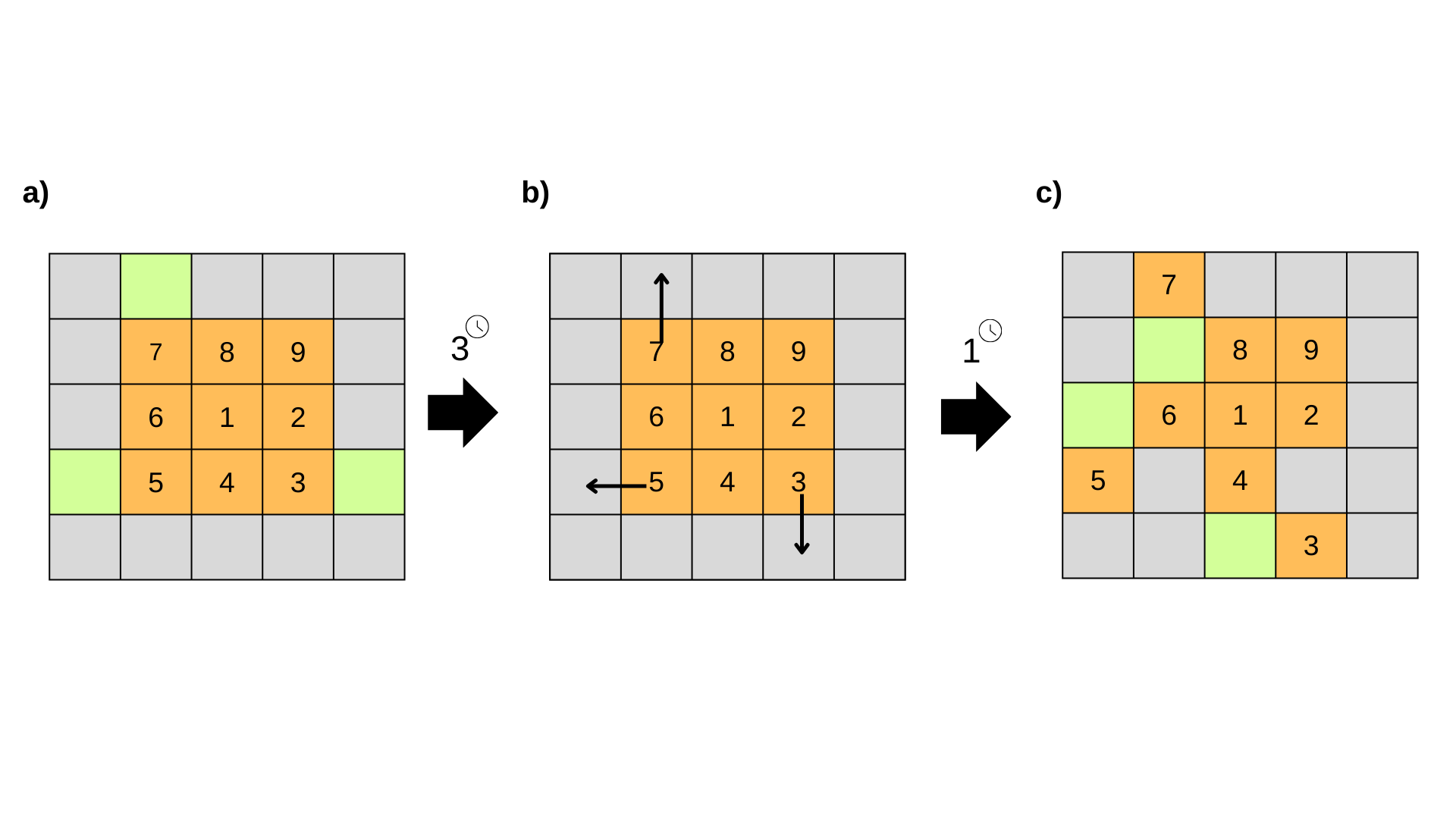}
  \caption{ Neighbour-dependent moves  inspect neighbouring qubits to determine the appropriate move operation. Before the CNOT operations, qubits labelled 3, 5, and 7 undergo a Hadamard operation simultaneously (\textbf{a}). Following this, the data qubits consult the circuit's directed acyclic graph (DAG) to determine the subsequent move operations. In this instance, the upcoming gates are three CNOTs applied to qubit pairs (3, 4), (5, 6), and (7, 8). To prepare for these interactions, the data qubits move to adjacent positions, enabling diagonal placement (\textbf{b}). The ancilla qubits positioned between each control–target pair are highlighted in neon. The Hadamard operation requires three timesteps, whereas each move operation requires one timestep.} 
    \label{fig:10}
\end{figure}

\subsection{Initial Mapping}

The initial mapping strategy involves assigning qubits based on the CNOT gates present in the circuit. For 2D condensed matter Hamiltonians, a mapping scheme is unnecessary due to the nearest-neighbour connectivity inherent in their Hamiltonians. For the remaining benchmark circuits, the program qubit pairs are placed onto the 2D grid in order of priority, based on the sequence of a specified number of CNOT (two-qubit) gates in the original circuit. The mapping algorithm uses the circuit’s \emph{directed acyclic graph} (DAG) as input. It constructs a list of CNOT qubit pairs ($cnot\_list$) sorted by frequency, where $k,v$ represents a program qubit $k$ and the set of qubits $v$ that it interacts with through CNOT operations. It then places these pairs onto the 2D grid (starting from cell $row,column=0,0$) in order of decreasing occurrence so that the most frequently interacting qubits are assigned first. Any program qubits left unmapped after the main procedure are assigned to the grid at random.

\subsection{Gate-Dependent Moves in the Neighbourhood}

  Gate-dependent moves constitute a form of look-ahead compilation, wherein, after applying an operation to a given data qubit, the compiler examines subsequent operations in the circuit before deciding the qubit’s movement. The heuristic takes two consecutive layers of the DAG of the quantum circuit at once, applies gates to qubits that satisfy the placement constraints (for example, CNOT needs a control and target qubit in a diagonal position with ancilla in between, as shown in Figure \ref{fig:5}).  It then determines the next set of move operations for the qubits based on the next layer of operations. An illustrative example is provided in Fig. \ref{fig:10}, where the first panel applies an H operation to three data qubits simultaneously (ancilla qubits highlighted in neon). Subsequent move decisions are then made to ensure that the following layer of operations, which involves a CNOT gate, respects the placement constraints. In the example shown, qubit labels 3, 5, and 7 move to positions diagonal to 4, 6, and 8, respectively, to enable a CNOT between them. 

 \begin{figure}  
  \centering
  \includegraphics[width=0.5\textwidth]{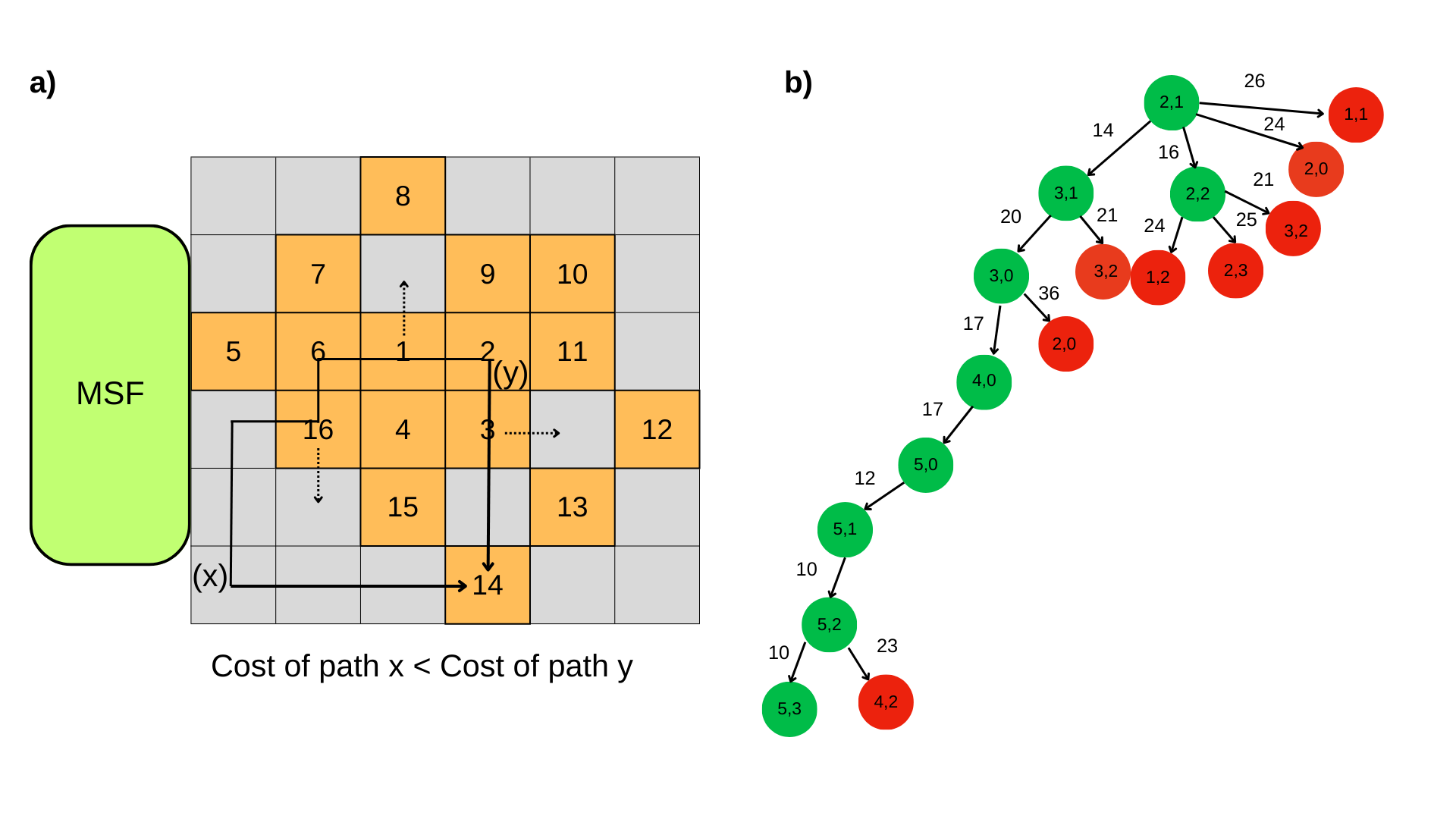}
  \caption{The starting position is qubit 6 at coordinates (2,1), and the target position is qubit 14 at (5,3). As illustrated in \textbf{(a)}, path (x) has consistently lower costs than path (y). Figure \textbf{(b)} shows the corresponding cost-accumulation graph, where the green cells indicate the optimal movement path. The red cells indicate the rejected cells. The cost of each movement is given on the edge and cells with the minimum cost are chosen.} 
 
    \label{fig:12}
\end{figure}

  \subsection{Find a Path using Dijkstra's Algorithm}

  In our construction, the move operations are essential to the compilation scheme, as data qubits need to be configured correctly. 
  Pathfinding is, therefore, the most important heuristic because of the large number of move operations that need to be made. We use Dijkstra's algorithm implemented with binary heap priority queue data structures to efficiently select the next qubit to process. The algorithm minimises a cost function to determine a near-optimal path (refer to Algorithm \ref{algo3}). Our pathfinding algorithm is used to route magic states, move data qubits for CNOTs, and find unoccupied cells for lattice surgery operations. The cost function in the algorithm is given by:
  
\begin{equation}
 {C}(a,b)= d(a,b) * step\_count
\end{equation}
  \noindent where d(a,b) is the distance between a and b. The $step\_count$ tracks the total number of steps in the path, including any penalties added when the path crosses an occupied data qubit. Accounting for occupied cells plays a crucial role because, in a dense 2D grid, the goal is to identify paths that minimise disturbance to other data qubits. The algorithm explores neighbouring cells and assigns a specific penalty cost to each potential move. Movement to an unoccupied cell incurs zero cost (except for the additional step), whereas moves over occupied cells accrue a penalty given by $2\times(count+weight)$. The weight can be varied depending upon the density of data qubits in the grid.  This procedure effectively creates a cost-accumulation graph that is dynamically updated each time a cell is moved. A simple example illustrating this process is provided in Fig. \ref{fig:12}. The path taken from (2,1) to (5,3) is selected by comparing the lowest cost among all the neighbouring cells for each step.  
  
\begin{algorithm} 
\caption{Path Finding: heap-based algorithm that selects the next grid position using the Manhattan distance and occupancy-dependent costs. \texttt{manhattan} calculates the Manhattan distance between two locations on a grid. The weight can be changed based on the density of the layout.}
\label{algo3}
\DontPrintSemicolon

\KwIn{$grid,\, start,\, end,\, weight$}
\KwOut{$path$ (list of coordinates), $cost$}

$heap \gets$ empty min-heap\;
$initial\_count \gets 0$\;
$initial\_dist \gets \texttt{manhattan}(start, end)$\;
$initial\_cost \gets initial\_dist \times initial\_count$\;
\BlankLine
\texttt{heap.push((initial\_cost,\, start,\, [start],\, initial\_count))}\;

\While{heap not empty}{
    $(cost,\, current,\, path,\, count) \gets$ \texttt{heap.pop\_min()}\;

    \If{$current = end$}{
        \Return{$(path,\, cost)$}\;
    }

    \ForEach{direction $(di,dj)$ in \{(0,1),(1,0),(0,-1),(-1,0)\}}{
        $neighbour \gets (current_i + di,\, current_j + dj)$\;

        \If{neighbour is outside grid or blocked}{
              \tcp*{skip invalid neighbour}
        }

        \If{$grid[neighbour] = 0$}{
            $step\_count \gets count + 1$\;
        }
        \Else{
            $step\_count \gets 2 \times (count + weight)$\;
        }

        $dist \gets \texttt{manhattan}(neighbour, end)$\;

        \If{$dist \neq 0$}{
            $new\_cost \gets step\_count + \texttt{manhattan}(start, neighbour)$\;
            $new\_path \gets path \cup \{neighbour\}$\;
            \texttt{heap.push((new\_cost,\, neighbour,\, new\_path,\, step\_count))}\;
        }
        \Else{
            \Return{$(path \cup \{neighbour\},\, cost)$}\;
        }
    }
}

\Return{(empty list, $\infty$)} \tcp*{no path found}
\end{algorithm}

\subsection{Space Search}
 \begin{figure}  
  \centering
  \includegraphics[width=0.5\textwidth]{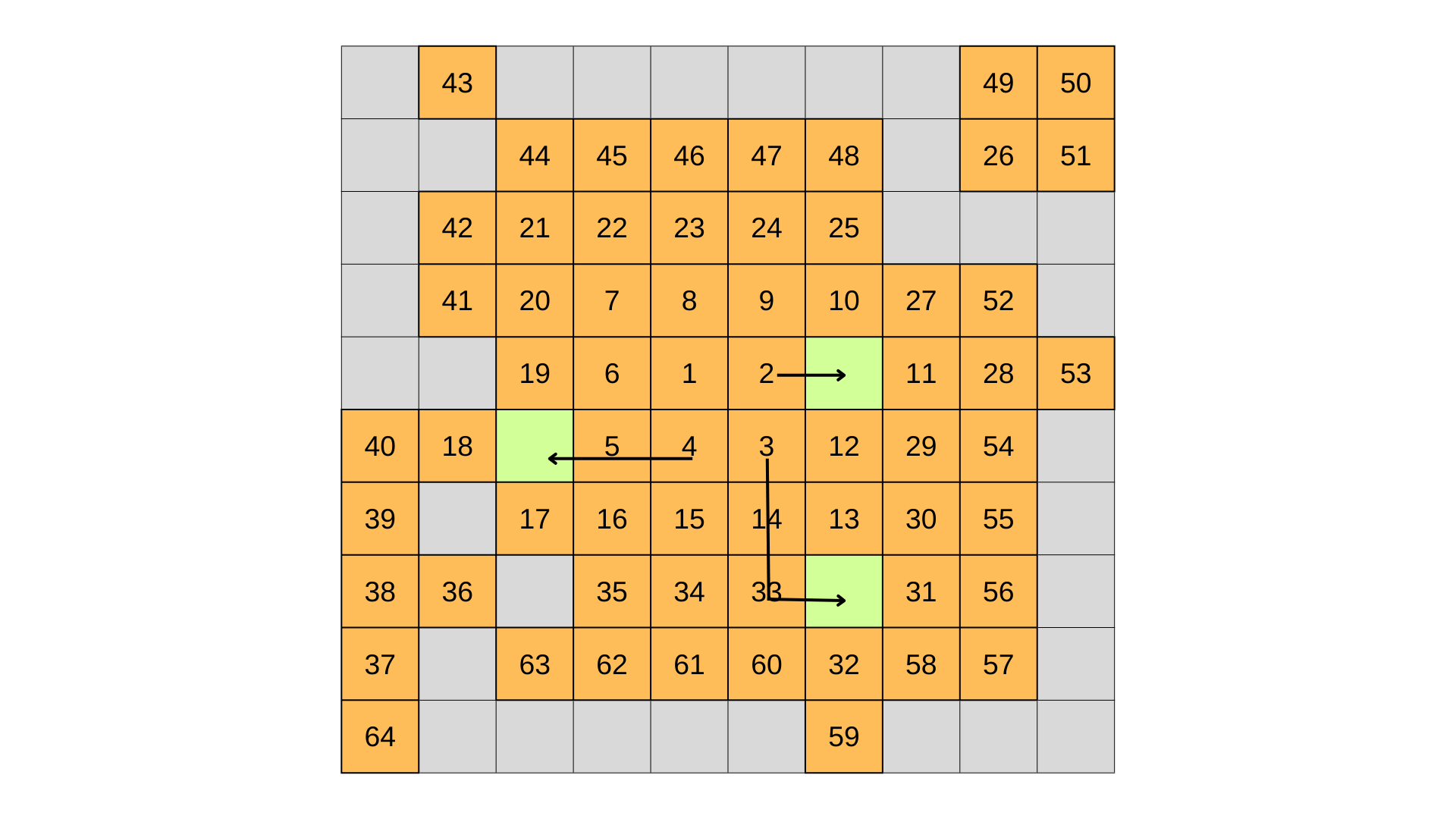}
  \caption{Space search identifies unoccupied cells around data qubits in order to create space within an ancilla-optimised layout while minimising the number of move operations. In the example shown, the data qubit labelled `1' undergoes a Hadamard operation, and three neighbouring data cells are available for movement. Since the Hadamard gate requires only a single ancilla, relocating the qubit labelled '2' is the most efficient option.} 
  \label{fig:11}
\end{figure}

 Space search is employed to determine an appropriate placement of ancilla for a specific gate operation. The algorithm takes as input the location of the target qubit and the operation to be applied. It then searches the 2D grid for the nearest unoccupied cell that can be used as a degree of freedom in the congested grid, moving the occupied cells one step at a time. This creates the necessary space for the operation, and the appropriate path is found using Dijkstra's algorithm with the preset cost function. Fig. \ref{fig:11} illustrates an example of a space search in a densely packed grid. The ancilla position that requires the fewest moves to clear is selected to minimise overall time overhead.

\subsection{Remove Redundant Moves}

  This layer of optimisation is carried out in the scheduling stage. The DAG, which is generated finally from the methods above, is used to find commuting move operations such that $U_{r_{i}\rightarrow r_{j}}^{\dagger}U_{r_{j}\rightarrow r_{i}}=I $, where $r_{i}, r_{j}$ correspond to the initial and final positions of the move operation. Since greedy heuristics make decisions based on the immediate achievable placement of data qubits, many redundant move operations can be identified and eliminated, thereby further reducing the overall execution time.


\begin{figure} 
  \centering
  \includegraphics[width=0.5\textwidth]{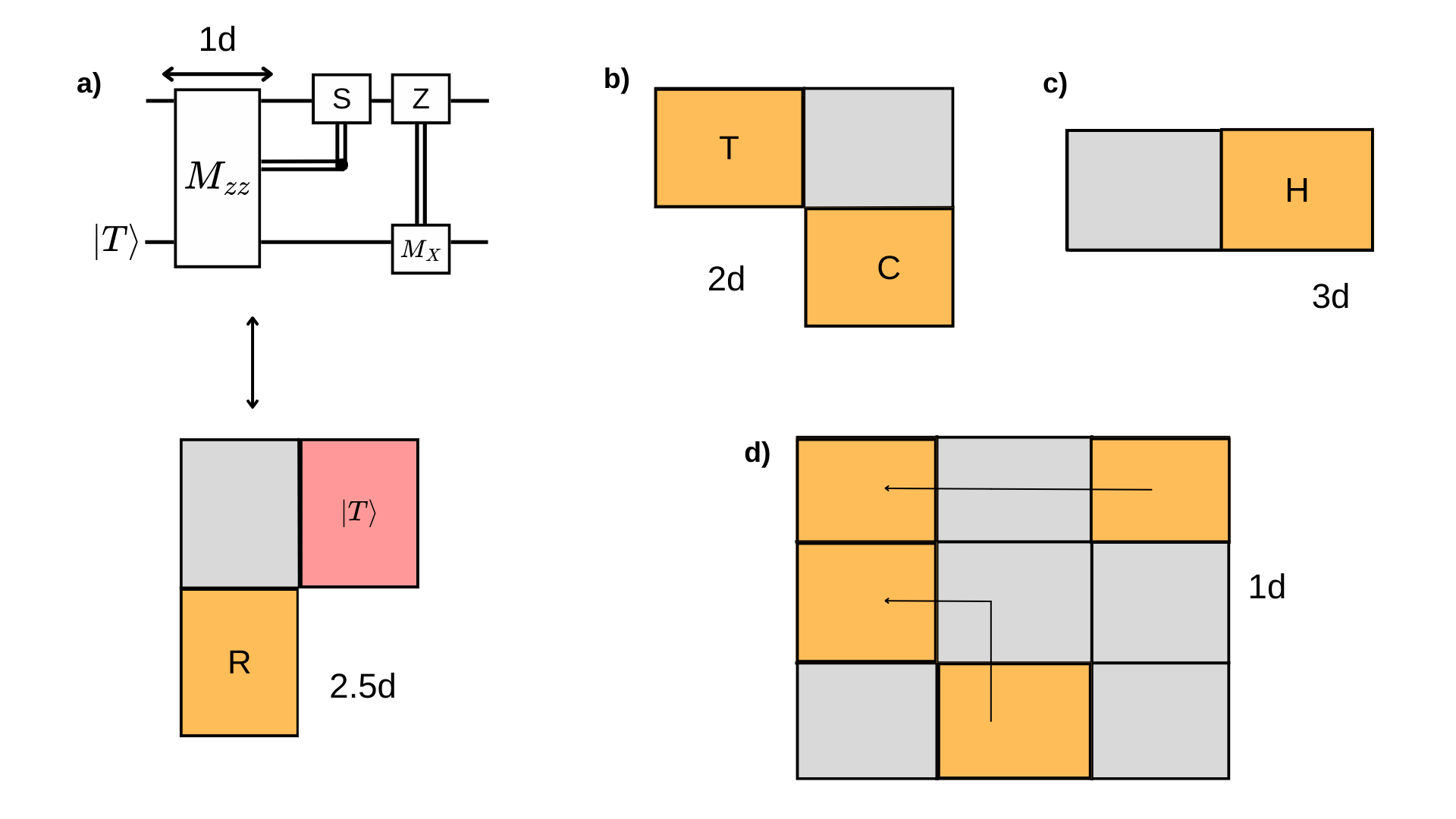}
  \caption{Lattice  surgery operations with their configurations and timesteps. Data qubits are given in orange, ancilla in grey and the magic state in pink. \textbf{(a)} T gate is applied in 2.5d timesteps, where a single $M_{zz}$ operation takes 1d timestep, followed by an S gate. (\textbf{b}) CNOT gate is applied using a control (C) and a target (T) qubit. (\textbf{c}) Hadamard assumes one ancilla and is performed in 3d timesteps. (\textbf{d}) Move operation takes 1d timestep.  } 
  \label{fig:3}  
\end{figure}
\vspace{-15pt}
\section{Evaluation}
\label{sec:evaluation}

 \textbf{Instruction set:} In Figure \ref{fig:3}, we describe the operations used by our compilation method and the time cost associated with them. We assume an instruction set where operations are allowed only between sets of logical qubits which are adjacent~\cite{10946814, PhysRevA.86.032324}. This instruction set choice eases implementation requirements for early FTQC systems. Following prior works, we mention the time cost per operation to be a multiple of the code distance (d)~\cite{Litinski2019gameofsurfacecodes, beverland2022assessingrequirementsscalepractical}. For example, a) T state consumption circuit shows an $M_{zz}$ operation along with S gate  \cite{Litinski2019gameofsurfacecodes}. The $M_{zz}$ operation in lattice surgery takes 1d code cycle, and the S gate takes another 1.5 d code cycles \cite{PhysRevResearch.6.013325}. (b) CNOT takes 2d timesteps because of $M_{zz}$ and $M_{xx}$ operations \cite{unknown} and  (c) Hadamard takes 3d timesteps \cite{PhysRevResearch.6.013325, PRXQuantum.4.020303}. Finally, (d) Move operation takes 1d timesteps \cite{PRXQuantum.3.020342}. 
 
 \textbf{Placement constraints:}  The placement of the data and ancilla qubits during an operation is crucial for the implementation. A logical qubit has both X and Z edges, and only relevant sides can participate in $M_{xx}$ and $M_{zz}$ operations. For example, the control and target qubits (as given in Figures \ref{fig:5} and \ref{fig:3}) should be diagonal with a neighbouring ancilla to perform both $M_{zz}$ and  $M_{xx}$ operations, respectively. In our scheme, we assume edge constraints that are synthesised into sequences of operations before invoking our compiler. This allows merge operations to take place only at relevant edges corresponding to Z and X syndromes. Therefore, $M_{ZZ}$ operations only occur vertically, whereas $ M_{XX}$ operations occur horizontally. 
 
 
\begin{table}[h] 
    \centering
     \caption{Gate counts for different benchmark circuits.}
     \label{tab:gate_counts}
     \renewcommand{\arraystretch}{0.9} 
    \begin{tabular}{|p{1.6cm}|p{4.2cm}|p{1.5cm}|}
        \hline
        \textbf{Model} & \textbf{Gate Count} & \textbf{Qubits} \\
        \hline
        \textbf{Ising 2D} & 
        $\mathbf{CNOT}: 360,\;\mathbf{R_Z}: 280,\; \mathbf{H}: 300$ & 100    \\
        \hline
        \textbf{Heisenberg 2D} & 
        $\mathbf{H}: 1440,\;\mathbf{CNOT}: 1080,\;\mathbf{R_Z}: 540,\;\mathbf{S}: 360,\;\mathbf{S^\dagger}: 360$ & 100   \\
        \hline
        \textbf{Fermi Hubbard 2D} & 
        $\mathbf{H}: 400,\;\mathbf{CNOT}: 300,\;\mathbf{S}: 100,\;\mathbf{S^{\dagger}}:100; \mathbf{R_z}: 150$ & 100     \\
        \hline
        \textbf{GHZ} & 
        $\mathbf{CNOT}: 254,\; \mathbf{R_Z}: 2,\;  \mathbf{SX}: 34,\;\mathbf{X}: 1$ & 255 \\
        \hline 
        \textbf{Adder} & 
        $\mathbf{R_Z}: 240,\;\mathbf{CNOT}: 195,\;  \mathbf{SX}: 48,\;\mathbf{X}: 13$ & 28\\
        \hline
        \textbf{Multiplier} & 
        $\mathbf{R_Z}: 300,\;\mathbf{CNOT}: 222,\;  \mathbf{SX}: 34,\;\mathbf{X}: 4$ & 15 \\
        \hline
    \end{tabular}
   
\end{table}

\textbf{Benchmarks:} We use a set of 18 quantum programs and compare our work with three state of the art FTQC compilation techniques. Our benchmarks are given in Table \ref{tab:gate_counts}. They include condensed matter Hamiltonians across system sizes ranging from 4 to 100 qubits (specifically 4, 16, 36, 64, and 100), though in the table we report only the maximum dimension. The circuits used for these condensed matter Hamiltonians are all single Trotter step implementations. 

\textbf{Lower bound calculation:} We consider a set of qubit layouts by varying the number of routing paths (as shown in Fig. \ref{fig:9}) and assess our results relative to a theoretical lower bound in execution time. Our lower bound (\textit{l}) is calculated by the formula:
\begin{equation}
    \textit{l}=\frac{ n_{\textit{T}}*t_{\textit{MSF}}}{n_{\textit{MSF}}}
    \label{eq1}
\end{equation}
\noindent where $n_{T}$ is the number of T states required in a quantum circuit, $t_{MSF}$ is the processing time to produce a single magic state and $n_{MSF}$ is the number of factories assumed in the compilation process. Since our focus is on early fault-tolerant systems, we do not assume an abundance of magic states. The magic state production time is taken to be 11d timesteps \cite{Litinski2019gameofsurfacecodes}. The lower bound describes an ideal scenario where the only bottleneck determining the execution time is the preset distillation processing time, ensuring no further room for optimisation in the compilation.

\section{Results}

\subsection{Execution Time Analysis with Respect to Lower Bounds}
\begin{figure} [t]
  \centering
  \includegraphics[width=0.5\textwidth]{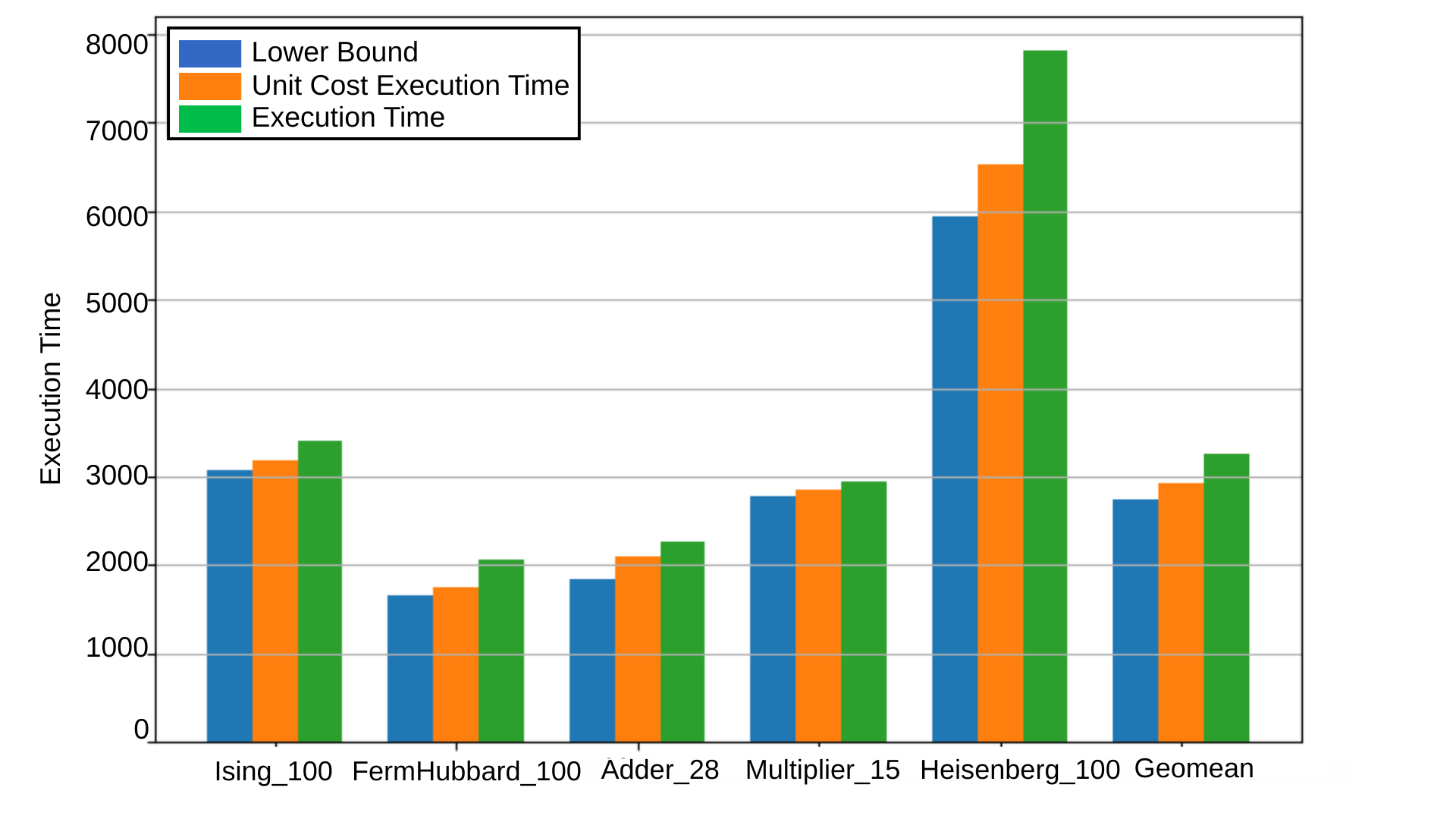}
 
    \caption{ Unit cost execution time is the total time calculated while assigning a unit cost to all the operations in the compiler scheme. Execution time refers to the actual time considering realistic operation latencies. On average, we record an increase of roughly 1.2$\times$ in the execution time compared to the lower bound.  } 
  \label{fig:14}
\end{figure}

  In Figure  \ref{fig:14}, the bar chart shows results of five benchmarks along with their geometric mean. The condensed matter circuits are all single Trotter step quantum implementations with 100 qubits. The results are presented for a specific qubit layout with $r=4$ (Figure \ref{fig:9}). This layout was selected as it yields promising results by incurring only minimal overhead in execution time. We compare our lower bound with both the execution time and the unit cost execution time. The execution time is the overall time taken to run the computation, while the unit cost execution time refers to the total time calculated by assigning a unit cost (1d timestep) to all operations in lattice surgery. This allows us to evaluate the proximity of our method to the ideal scenario where the only bottleneck in the circuit is the magic state processing time. In the case of condensed matter circuits, the unit cost execution time for Ising and Fermi Hubbard circuits is only  1.1$\times$-1.2$\times$  of the lower bound, while it is slightly higher (1.3$\times$) for the Heisenberg circuit. The total execution time (i.e., the time calculated by assuming realistic operation latency) is increased by a factor of 1.2 to 1.4 for the three circuits. In the multiplier circuit, the execution time is only 1.06$\times$ the lower bound. Overall, our compiler offers execution times that are very close to the lower bound.

 \begin{figure}[t] 
  \centering
  \includegraphics[width=0.5\textwidth]{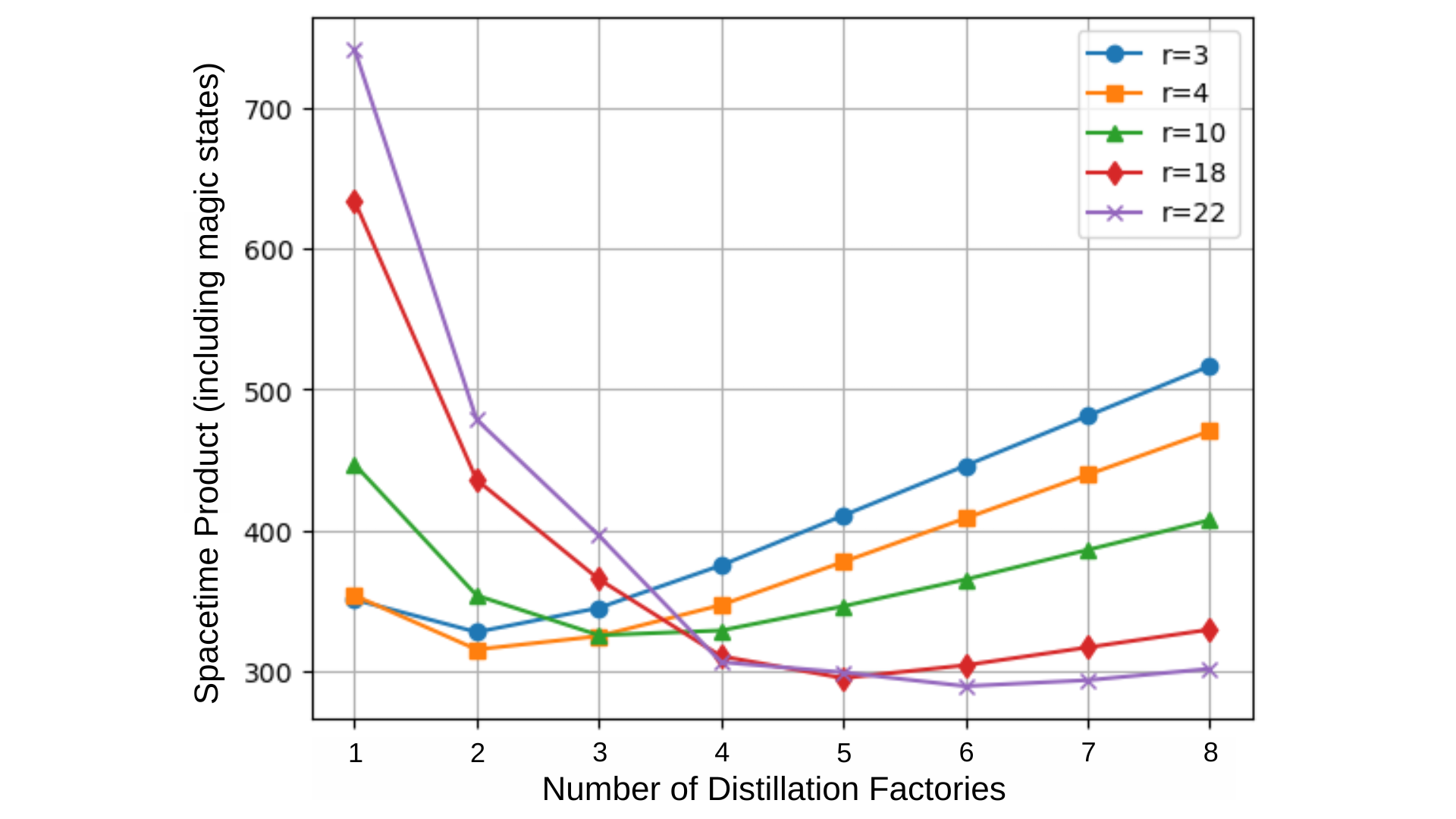}
  \includegraphics[width=0.5\textwidth]{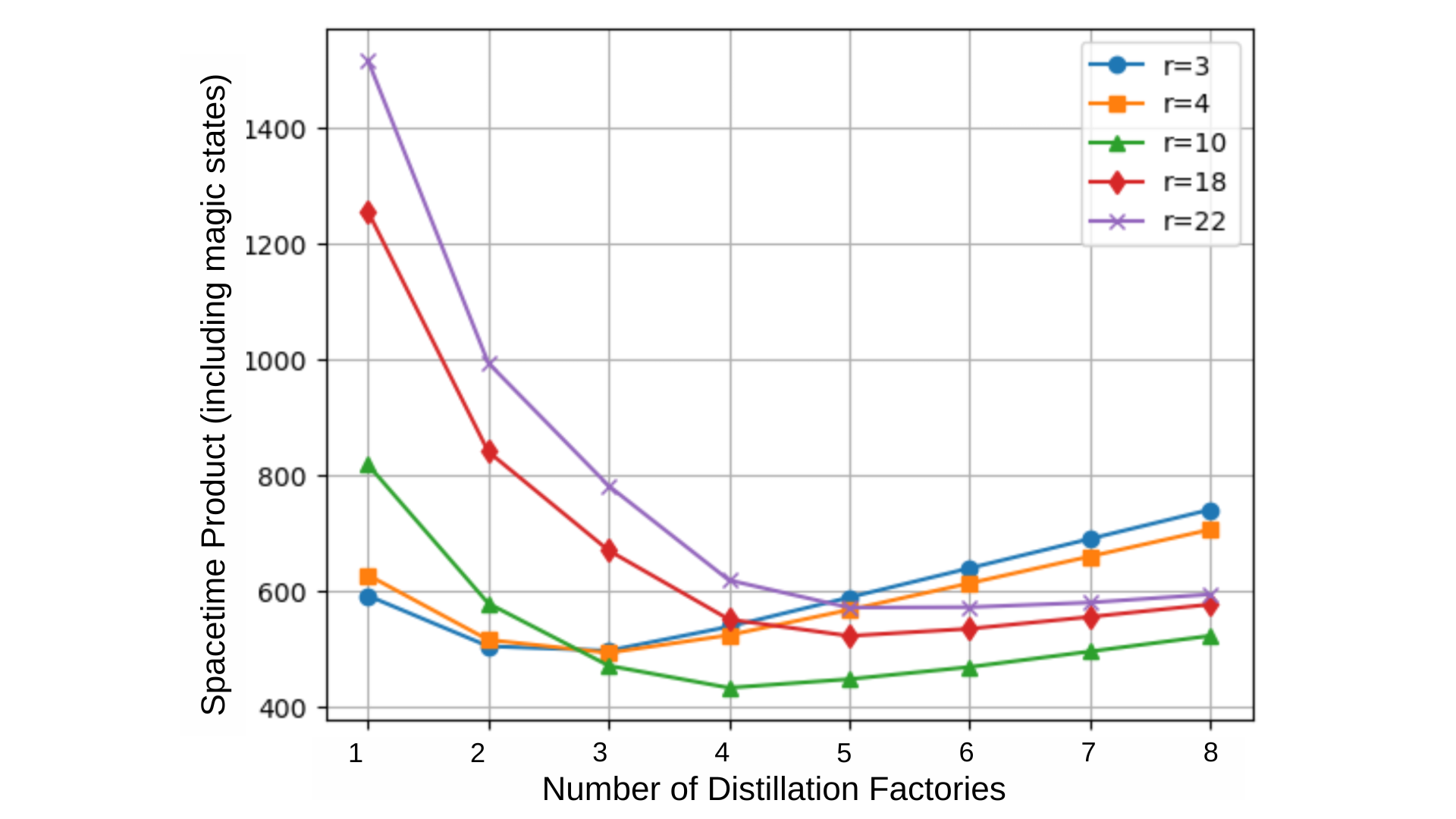}
  \includegraphics[width=0.5\textwidth]{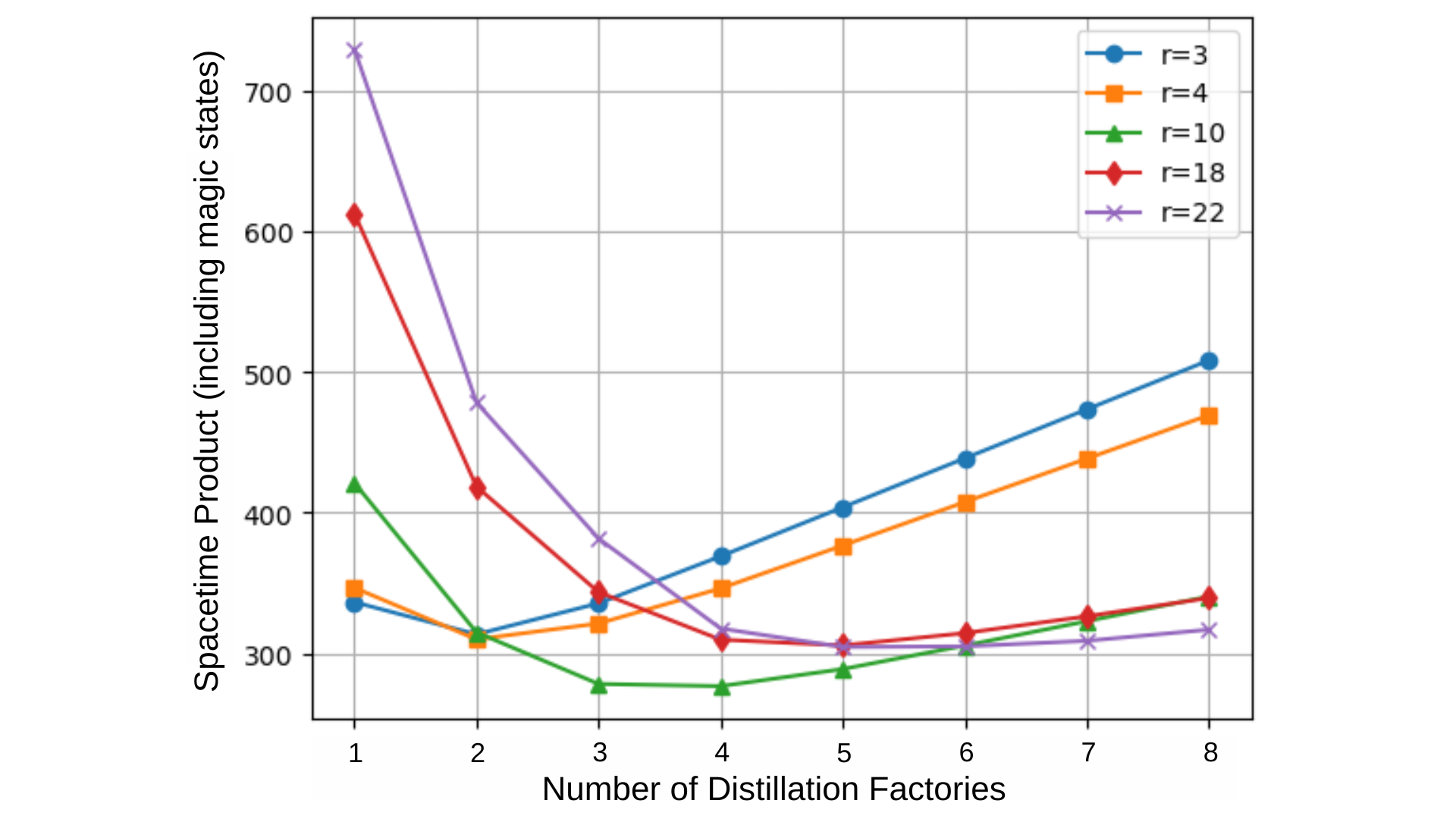}
 \caption{Spacetime volume (including magic state factory qubit requirements) per operation versus the number of distillation factories, with routing paths varied as specified in Figure~\ref{fig:9}. Top: 10x10 Fermi–Hubbard model. Middle: 10x10 Ising model.  Bottom: 10×10 Heisenberg Model. The parameter `r' denotes the number of routing paths. The trend shows that more routing paths are needed to obtain the benefits of a greater factory count. For $r= 3 $, two factories are optimal, but for $r=22 $, five factories give the lowest spacetime volume. }
  \label{fig:35,36}
\end{figure}
\subsection{Distillation-Adaptive Routing Path Allocation}
Figure  \ref{fig:35,36} has three plots depicting the spacetime product v/s the number of distillation factories while varying the routing paths. In these plots, the spacetime product includes distillation qubits and accounts for the total spacetime volume of computation. Each curve in the plot provides us with an optimal number of distillation factories for a layout with a given number of routing paths. We observe that the curves are U-shaped, owing to the trade-off between the number of qubits and execution time. The number of distillation factories is directly proportional to the number of qubits, and we observe that with too few qubits (fewer factories), runtime dominates. As we increase the factory count, time decreases, but qubit overhead dominates. The optimum lies in between, where execution time and qubit count balance out.  For example, in the Fermi-Hubbard 2D case (top), provisioning two distillation factories for layouts with 3-4 routing paths gives the optimal spacetime volume. When the routing paths (and qubits) are increased,  5-6 distillation factories are optimum ($r=18,22$). As the number of routing paths in the grid increases, greater flexibility is available for routing magic states, thereby making the impact of multiple distillation factories more pronounced. We also observe a marked shift in the spacetime volume when increasing the number of factories from one to eight. For the Heisenberg circuit (bottom), when operating with a single magic state factory, the spacetime volume for $r=22$ is approximately 2.16$\times$ that at $r=3$. In contrast, with eight factories, this relationship is inverted: the spacetime volume at $ r=22$ is reduced to about 0.62$\times$ the corresponding value at $r=3$. 

Moreover, our compiler offers the ability to determine the optimal combination of factory count and routing paths for each  application. For the Ising and Heisenberg circuits, an $r=10$ layout yields the lowest spacetime volume, whereas for the Fermi–Hubbard circuit the minimum occurs only when the number of routing paths exceeds 18. This variation occurs due to the differences in the instruction mix across applications. More magic states require more routing paths to relieve routing congestion and achieve low space-time volume.  By optimising these choices and reducing resource requirements, our work has the potential to enable early FTQC applications sooner.

 \begin{figure} [t]
  \centering
  \includegraphics[width=\linewidth]{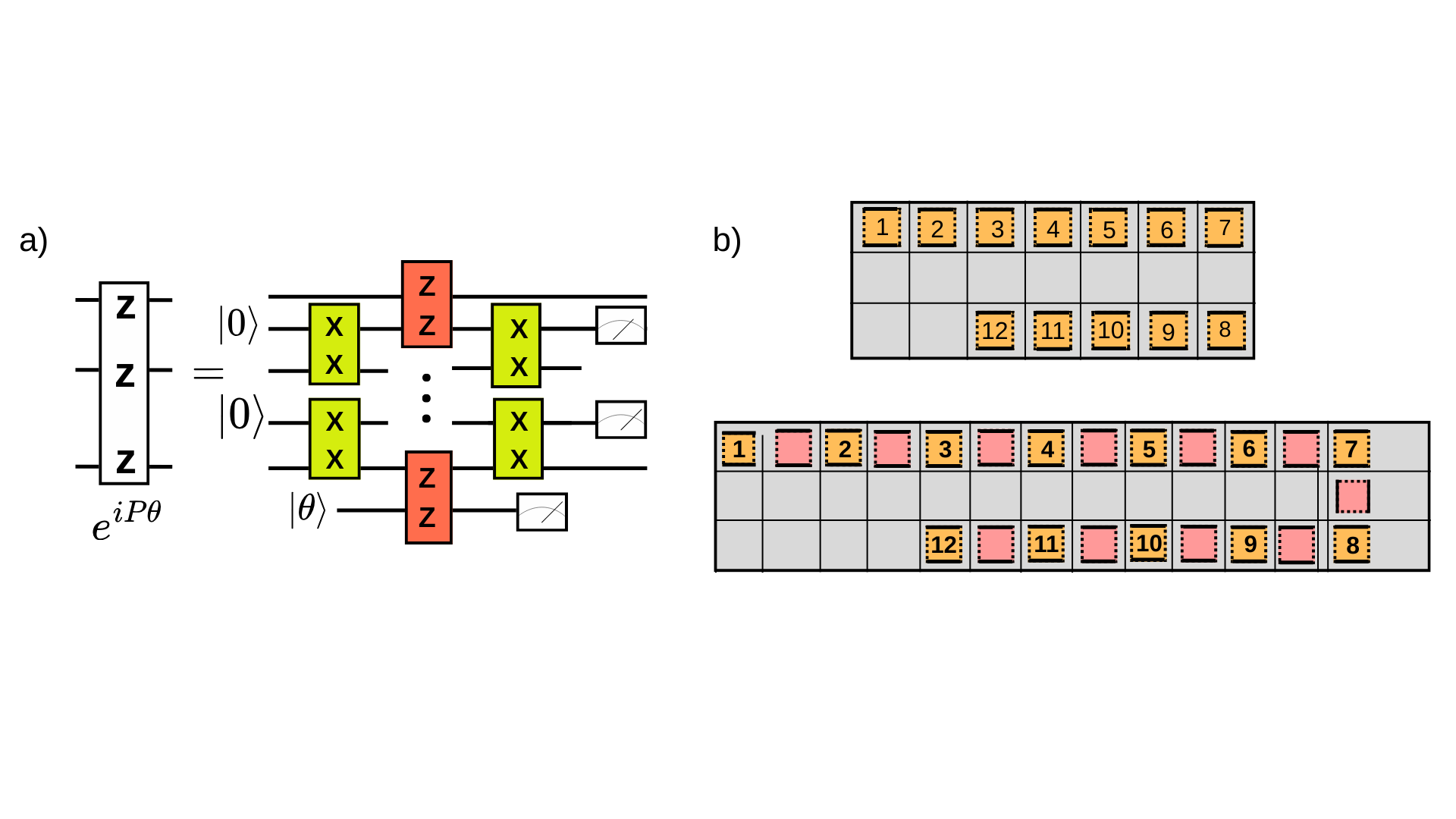}

  \caption{(\textbf{a}) Constant depth decomposition of Pauli-Product Rotations ($e^{iP\theta}$) using n ancillary qubits. (\textbf{b}) To run the same operations in a compact layout, new ancillary spaces are added (in pink). The bus qubits (grey) are then used as routing paths. The total number of qubits increases to 3n+3. } 
  \label{fig:24}
\end{figure}

\subsection{Comparison with Game of Surface Codes}
We evaluate our scheme against selected existing methods in the literature, comparing both qubit count and execution time. We draw comparisons with techniques from \cite{Litinski2019gameofsurfacecodes,10946814,10.1145/3720416}. 
In Figure  \ref{fig:15,20,21}, we compare the results from \cite{Litinski2019gameofsurfacecodes}, using its compact and fast block layouts. The execution time of Pauli Product Rotations (PPRs) is determined by first decomposing the PPRs into two-body XX and ZZ interactions, as given in \cite{Moflic:2024xhe}.  PPRs are large logical operations that comprise hundreds of qubits in a layout. Such operations are difficult to implement and are decomposed into nearest-neighbour operations assuming hardware connectivities. The decomposition described in \cite{Moflic:2024xhe} incurs an additional constant depth latency and requires twice the number of ancillary qubits for implementation (shown in Figure \ref{fig:24}(a)). To understand the realistic implementation of these PPRs, we introduce ancillary qubits in the layout design and calculate the spacetime cost accordingly. In the case of the compact block layout (Figure \ref{fig:24}(b)), this decomposition increases the number of qubits from 1.5n+3 to 3n+3, where n is the number of data qubits. Similarly, for intermediate and fast blocks, the total qubits in these layouts increase to 4n and 4n+6  when used for realistic implementations.
 \begin{figure}[t] 
  \centering
  \includegraphics[width=0.5\textwidth]{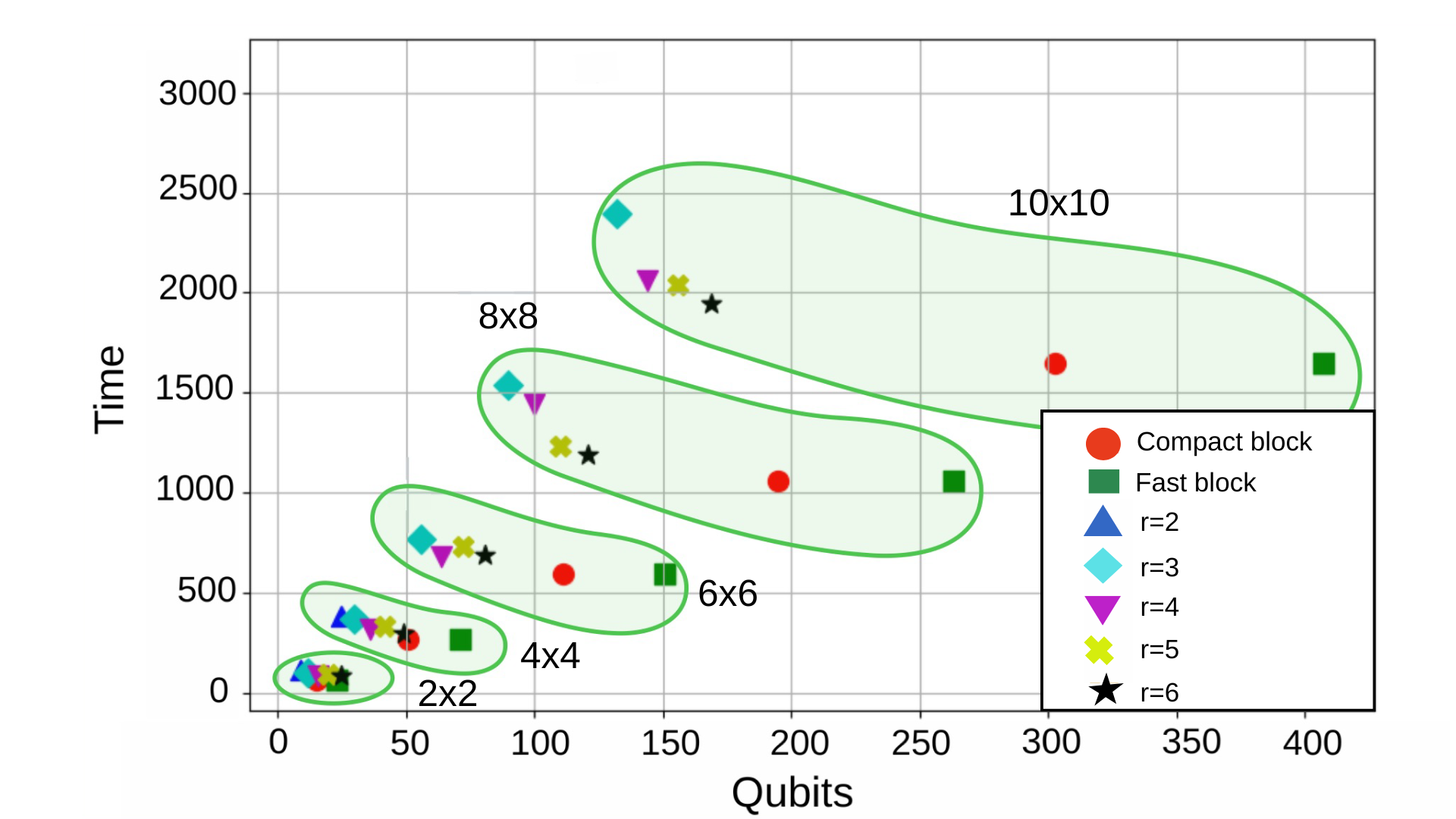}
  \includegraphics[width=0.5\textwidth]{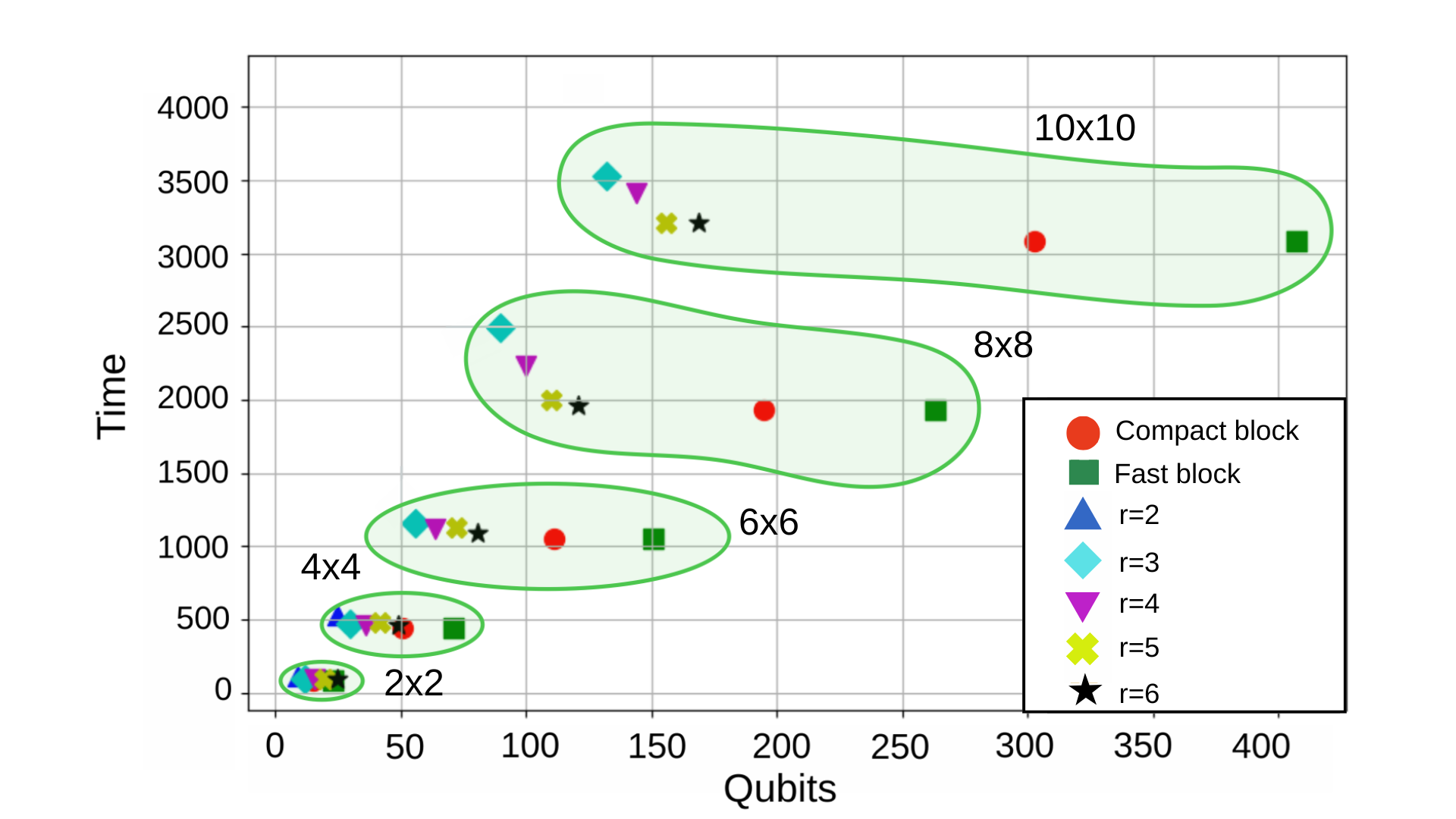}
  \includegraphics[width=0.5\textwidth]{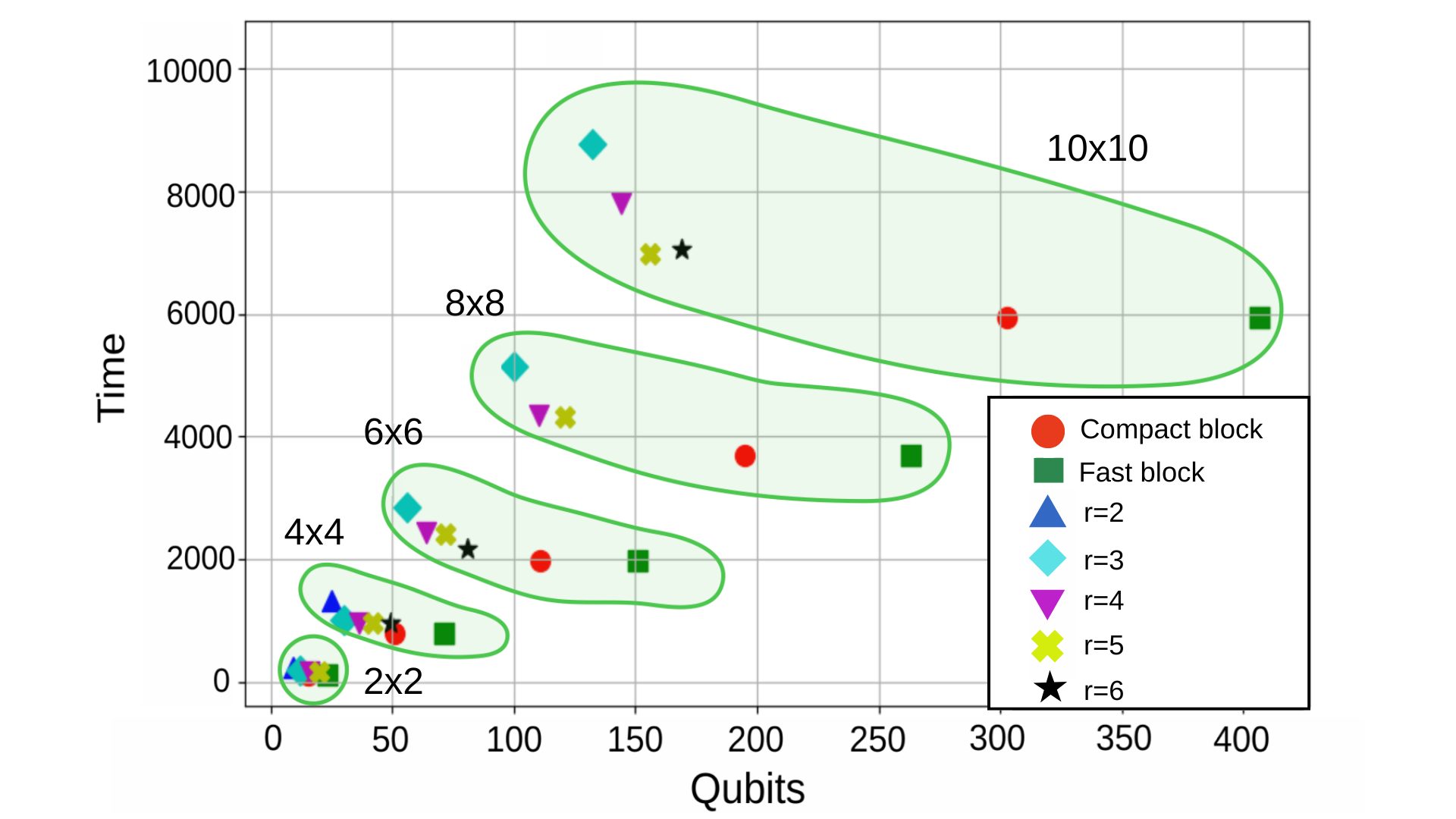}
  \caption{Execution time v/s qubit plots with one distillation factory and  single Trotter step: (Top) Fermi–Hubbard, (Middle) Ising, and (Bottom) Heisenberg model circuits, evaluated across problem sizes with routing paths as specified in Figure~\ref{fig:9}. The green circles indicate the problem sizes. Our results are compared with the compact and fast block layouts of \cite{Litinski2019gameofsurfacecodes}, under the assumption of constant-depth decomposition for large Pauli product rotations.} 
  \label{fig:15,20,21}
\end{figure}

 We note that the PPRs generated by moving the Cliffords using \cite{PennyLane-pauli-product-measurement} are not all $P= Z^{\otimes n}$ for unitary gates of the form  $e^{iP\theta}$ . Condensed matter Hamiltonians generate PPRs where   $P =X^{\otimes n} , Y^{\otimes n}$ and $P =Z \otimes I\ldots \otimes Z  $ among others. In particular, for rotations of the form $e^{i\theta  Z \otimes I\ldots \otimes Z }$, the presence of identity operators between the non-trivial terms would, in practice, require additional SWAP operations (and qubits), effectively mapping the gate onto a next-nearest-neighbour architecture. For the sake of generality, however, we do not account for this overhead and instead assume an equal number of timesteps for all multi-qubit Pauli-product rotations of the form $e^{iP\theta}$. 
 
 While the decomposition significantly increases the number of total qubits for the compact, intermediate and fast blocks, it doesn't affect the total time. This is because the magic state production time is greater than the operation execution time.  Nevertheless, the latency of such operations could become significant in cases where the magic state processing time is improved, more than one distillation factory is used or when SWAP gates are included to implement $P=e^{iZ\otimes I \ldots\otimes Z}$ operations.  However, in this case, the execution time of the PPR approach in all three layouts coincides with the lower bound presented in Equation \ref{eq1}.

  \begin{figure}[t] 
  \centering
  \includegraphics[width=0.5\textwidth]{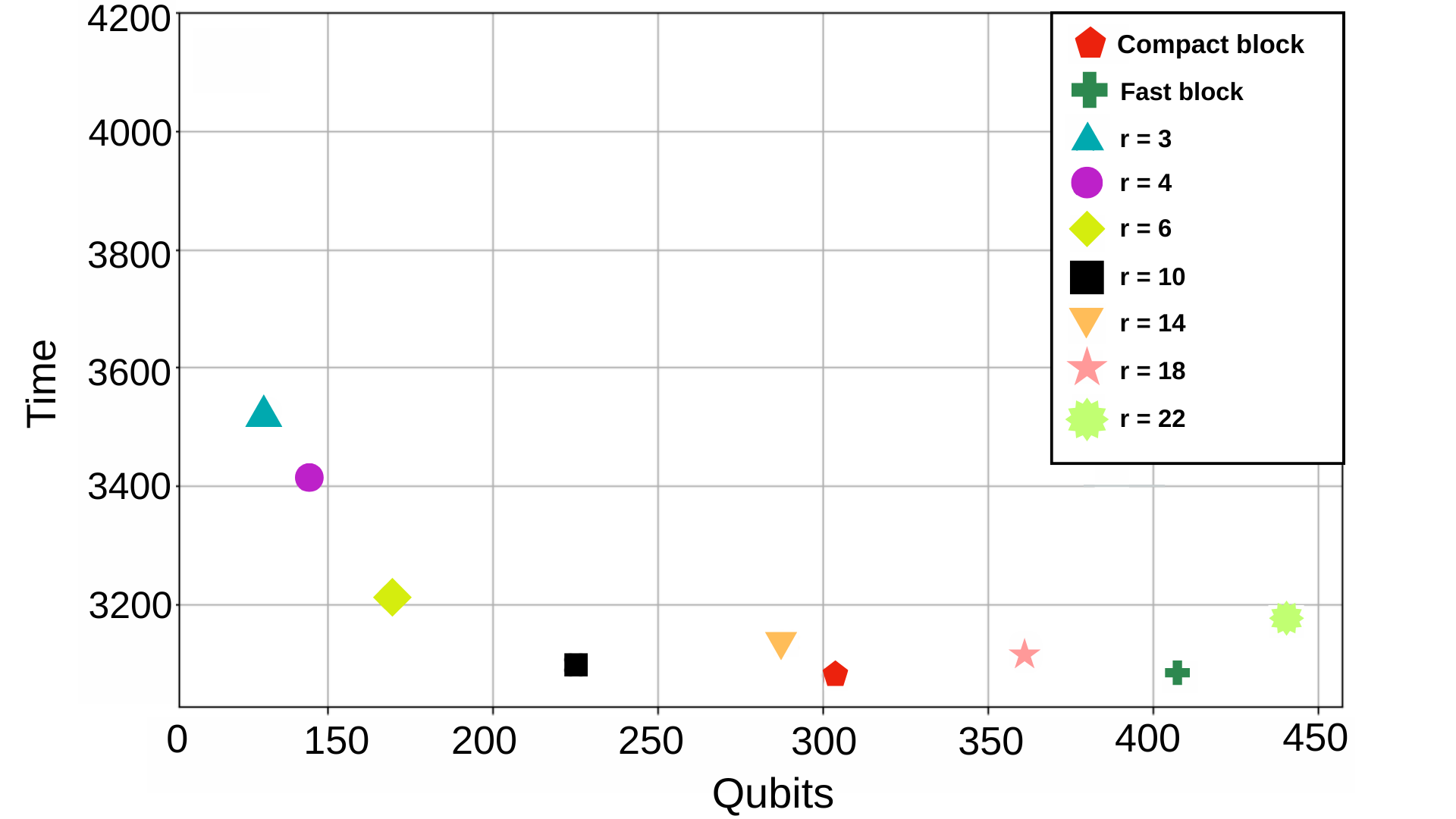}
  \includegraphics[width=0.5\textwidth]{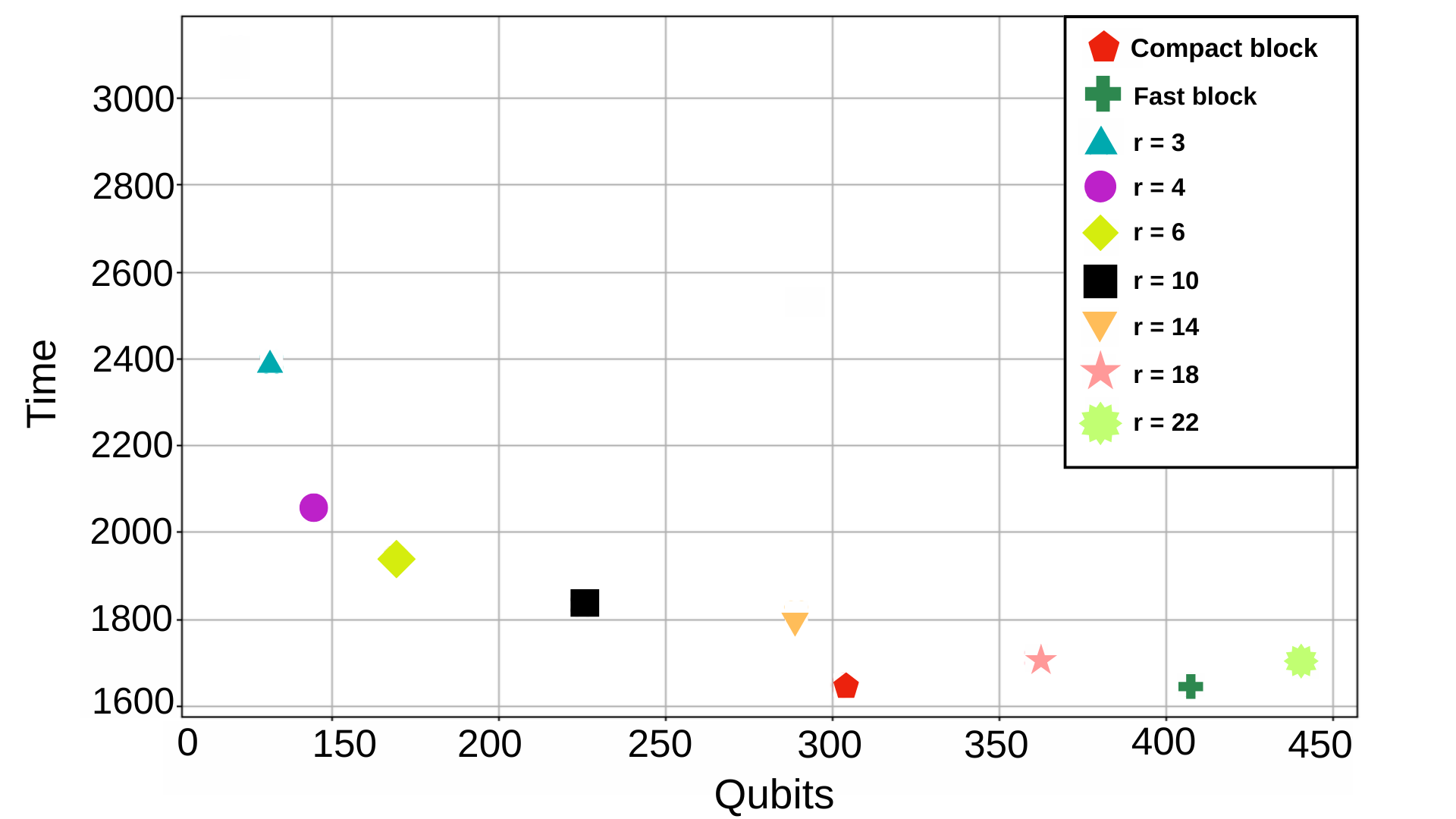}
   
  \caption{Execution time v/s qubit plots with one distillation factory and single Trotter step: (Top) Ising, (Bottom) Fermi–Hubbard model circuits, evaluated for 10x10 Hamiltonians with routing paths as specified in Figure~\ref{fig:9}. The results are compared against the fast and compact blocks. We obtain optimal results at r=10 with an execution time overhead of 1.04$\times$ for the Ising circuit and at r=6 with an overhead of 1.1x for the Fermi–Hubbard circuit. } 
  \label{fig:22,23}

\end{figure}

\begin{table}[t]
\centering
\setlength{\tabcolsep}{3pt} 
\renewcommand{\arraystretch}{1.1} 
\caption{Accuracy ($1-\epsilon$) comparison on benchmarks of different sizes. \cite{Litinski2019gameofsurfacecodes}'s columns use compact block, whereas our approach uses layout with $r=4$.   }
\label{tab:model-results}
\begin{tabular}{|l|c|c|c|c|}
\hline
 {} & \multicolumn{1}{c}{$d=11$} & { } & \multicolumn{1}{c}{$d=13$} &  {} \\

\hline
Model &  [27] & Our Approach &   [27] & Our Approach \\
\hline
Ising Model (144)    & 0.495  & 0.976 & 0.949 & 0.997 \\
\hline
Heisenberg (144)     & 0.0214 & 0.896 & 0.902 & 0.989 \\
\hline
Fermi Hubbard (144)  & 0.733  & 0.975 & 0.973 & 0.997 \\
\hline
Ising Model (100)    & 0.501  & 0.985 & 0.950 & 0.998 \\
\hline
Heisenberg (100)     & 0.037  & 0.954 & 0.903 & 0.995 \\
\hline
Fermi Hubbard (100)  & 0.732  & 0.986 & 0.985 & 0.998 \\
\hline
Ising Model (64)     & 0.872  & 0.991 & 0.987 & 0.999 \\
\hline
Heisenberg (64)      & 0.755  & 0.972 & 0.975 & 0.997 \\
\hline
Fermi Hubbard (64)   & 0.930  & 0.992 & 0.993 & 0.999 \\
\hline
\end{tabular}
\end{table}

 As shown in Figure \ref{fig:15,20,21}, we run our compilation technique on single Trotter step circuits of Fermi-Hubbard 2D, Ising 2D and Heisenberg 2D circuits. The circuit size is increased from 4 to 100 qubits, and the number of distillation factories is set to 1 (at 11d timesteps). We compare the resulting execution time (y-axis) from our techniques as a function of the qubit count (x-axis) with both compact and fast block layouts. The legends represent the number of routing paths in the layout. We observe that layouts with $r = 5$ and $6$ are the most optimal, achieving minimal execution time increase as compared to  compact and fast blocks, while offering a large reduction in qubit count. For 2D Ising programs, we see that the modified intermediate and fast blocks from \cite{Litinski2019gameofsurfacecodes}'s scheme have more than twice the number of qubits used in our approaches, with an average reduction of only about 20\% in overall time.  For the 100-qubit Fermi–Hubbard, Ising, and Heisenberg circuits, our best-case results achieve execution time overheads of  roughly 1.22$\times$, 1.04$\times$, and  1.17$\times$, respectively, while reducing qubit requirements by 53\%. This reduces the qubit count close to the optimal level, yielding a substantial reduction that brings the data-to-ancilla qubit ratio to approximately 3:1.

 \begin{figure}[t] 
  \centering
  \includegraphics[width=0.5\textwidth]{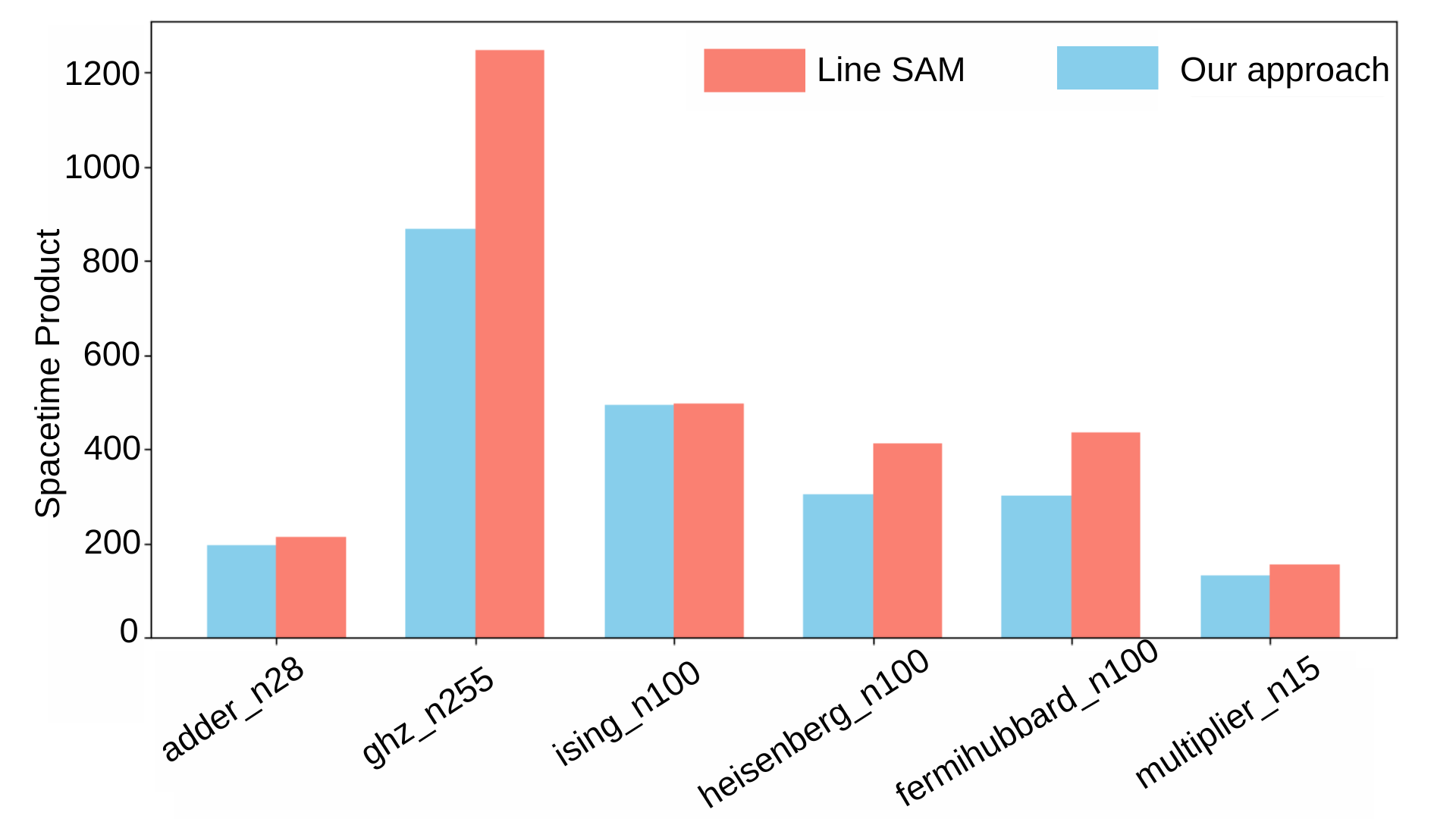}
  \includegraphics[width=0.5\textwidth]{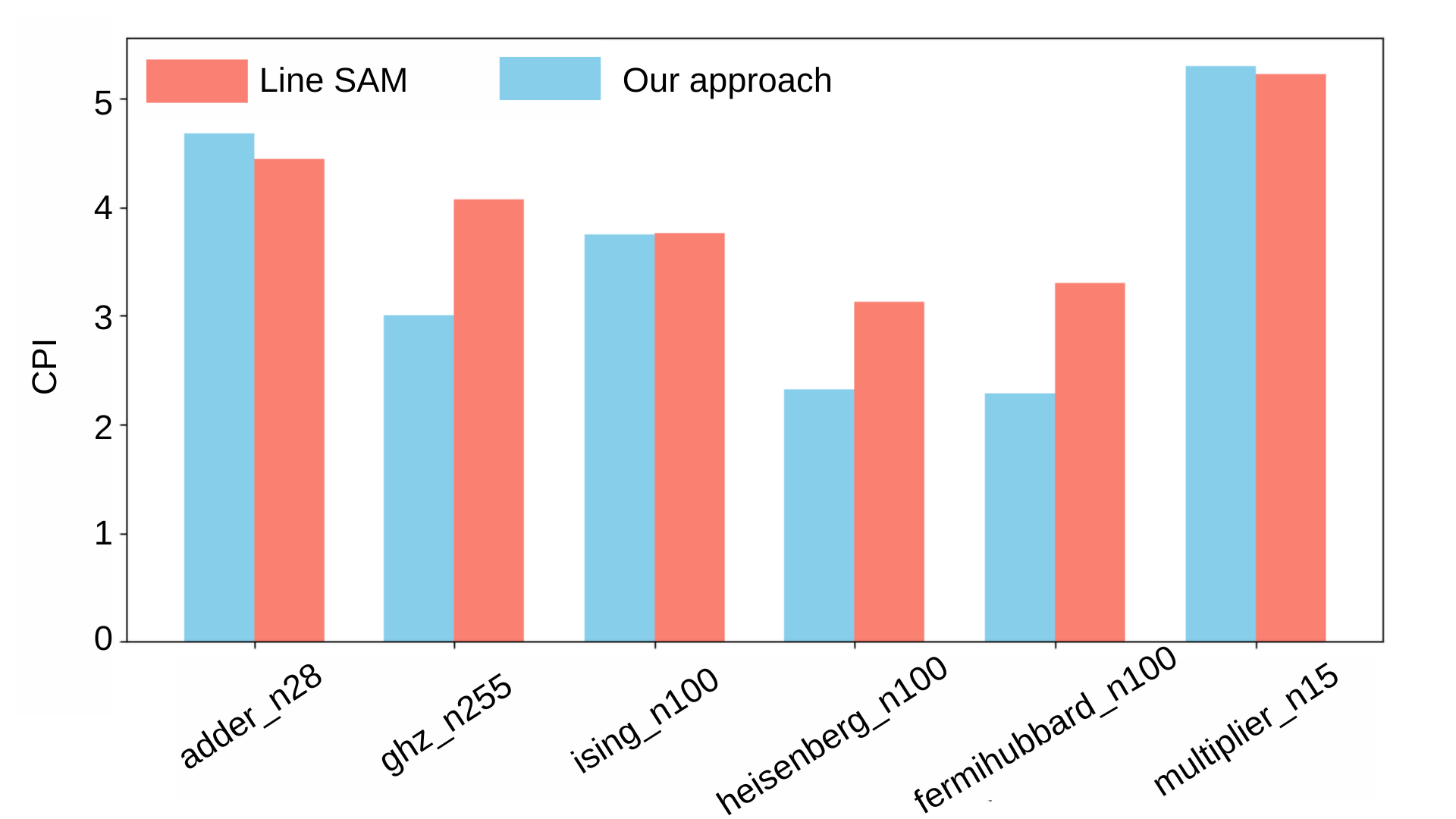}
  \includegraphics[width=0.5\textwidth]{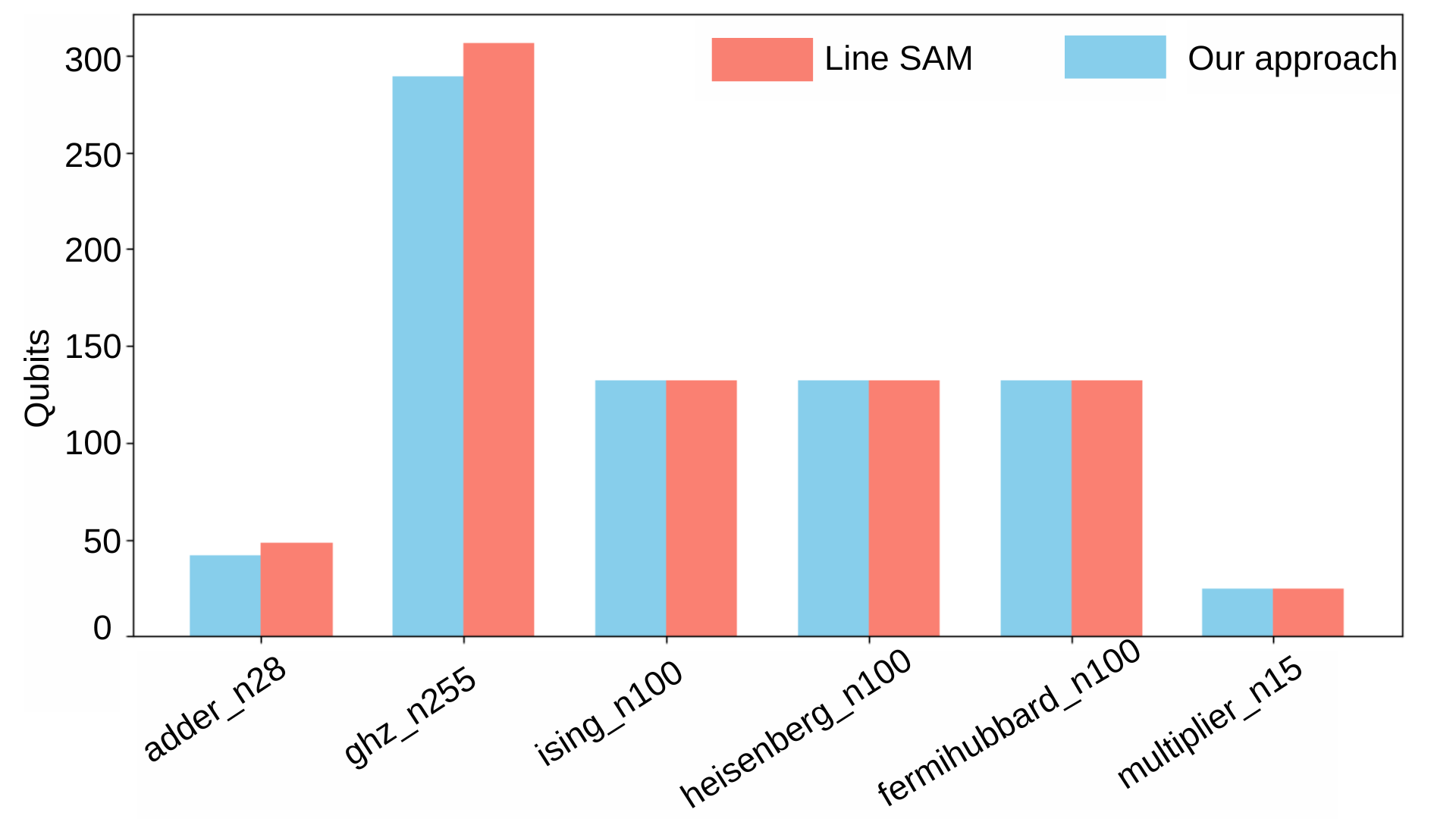}
  \caption{ Comparison with the Line-SAM method of \cite{10946814} with a single distillation factory. The top plot compares the spacetime volume of both approaches across the benchmarks listed in Table~\ref{tab:gate_counts}. The middle and bottom plots present the corresponding qubit counts and execution times, respectively. We get, on an average, a 20\% decrease in spacetime volume across all benchmarks.} 
  \label{fig:16}
\end{figure}

 In Figure \ref{fig:22,23}, we generate the execution time v/s qubit plots for 10x10 qubit Ising (top) and Fermi-Hubbard (bottom) circuits by varying the routing paths. We begin with 2 routing paths that have bus qubits aligned across 2 edges of the grid. The maximum number of routing paths in our layouts is 2n+2, where n is the number of data qubits. In this case, the maximum number of routing paths is 22, which is 3$\times$ the number of ancillary qubits per data qubit. Results show that the optimal range to implement these circuits is between 4 and 6 routing paths, which corresponds to 144-169 qubits. The overall time in compact and fast blocks is the same as the lower bound and is the lowest in the curve. Our method, with the same number of qubits as the compact and fast block ($\sim$400), has an overall time 1.03$\times$ more than the lower bound. We note that our technique is able to generate better, near-optimal results in conditions with a constrained set of resources, i.e., less provisioning of qubits.   

To provide an error analysis of our approach with \cite{Litinski2019gameofsurfacecodes}'s scheme on the basis of accuracy, we compute the overall fidelity of our implementation. We assume the physical qubit error rate (\textit{p}) to be 0.1$\%$ \cite{beverland2022assessingrequirementsscalepractical}. The logical error probability can be calculated as $P_{L}=0.03\times \frac{p}{0.01}^{(\frac{d+1}{2})} $. The maximum error ($\epsilon$) of a computation is given by \textit{spacetime volume $\times$ $P_{L}$}. Therefore, the overall fidelity can be given by  1 - $\epsilon$. The spacetime volume for the techniques in \cite{Litinski2019gameofsurfacecodes} utilises all qubits in the grid at all timesteps, resulting in a large spacetime volume. Through our compiler optimisations, we do not utilise all ancillary qubits in the grid at all timesteps, reducing the space time volume and increasing overall fidelity. For a given code distance in the architecture, we compute the overall fidelity in Table \ref{tab:model-results}. We consider $d=11$ and $ 13$, and compare our results (for $r=4$) with the compact block layout on different benchmarks. For the early FTQC regime, we are interested in lower code distances. For $n=100$, At $d=11$, we get a 1.3-25X improvement in error rate, with the 25X case occurring when the baseline suffers from very poor fidelity, but our approach has a high fidelity. When a higher code distance is used (which would result in a 40$\%$ larger quantum computer), our work still offers better fidelities than the baseline, but the improvement factor reduces.

 \begin{figure} 
  \centering
  \includegraphics[width=\linewidth]{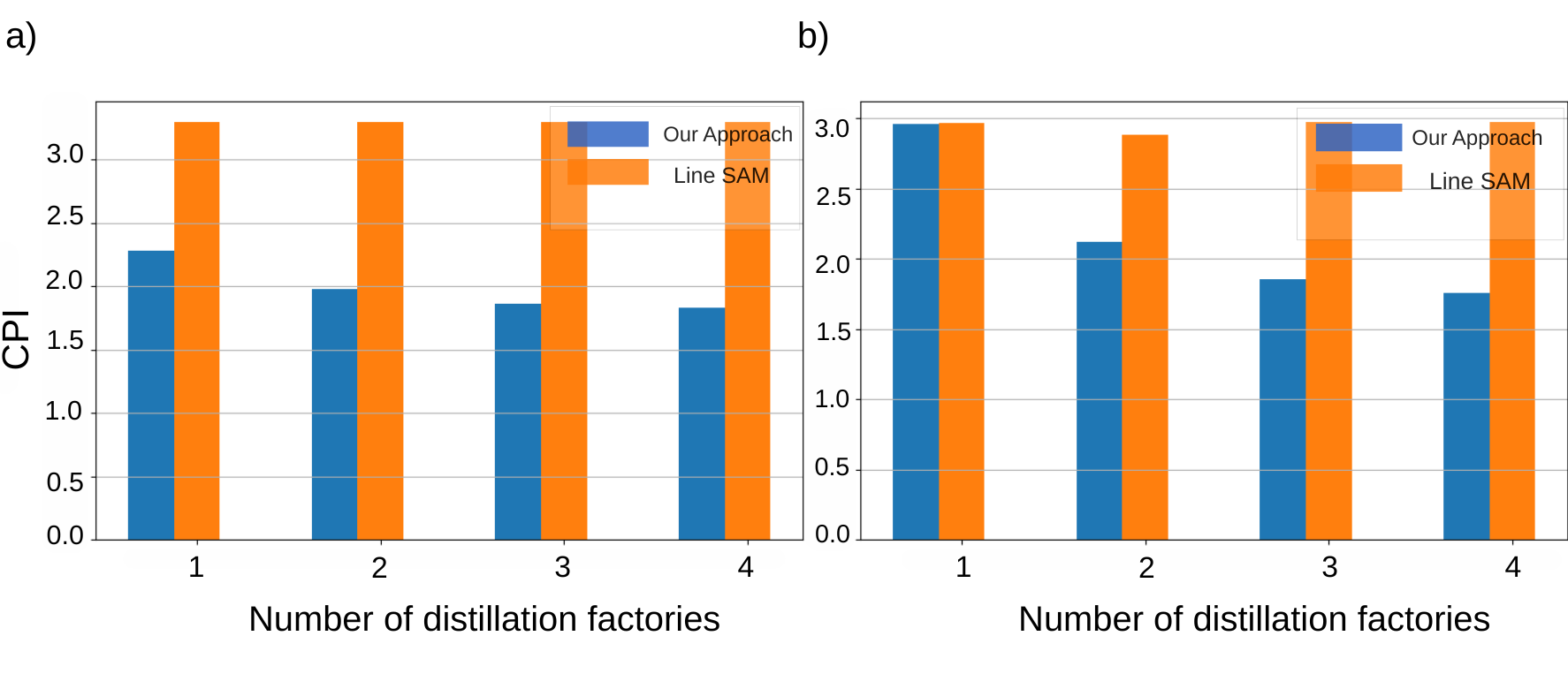}

 \includegraphics[width=\linewidth]{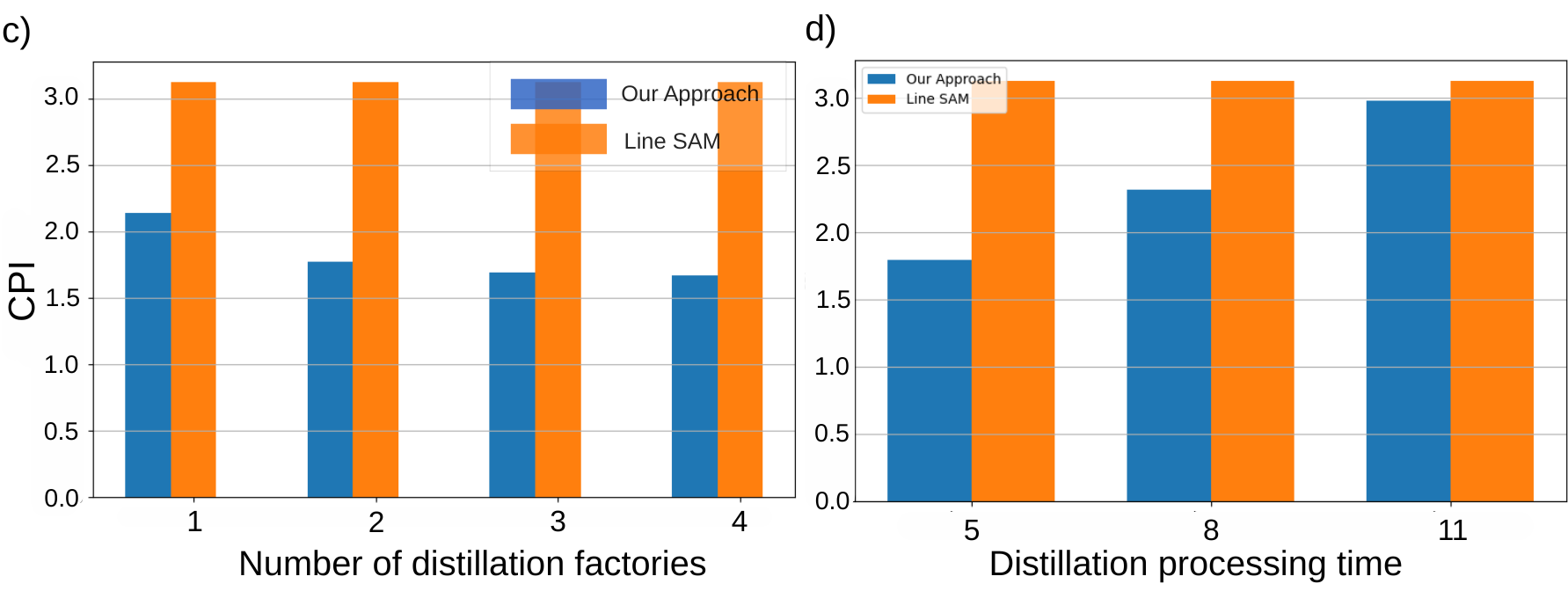}
  \caption{Comparison with the Line-SAM method of \cite{10946814}.   (\textbf{a–c}) present CPI results for 10×10 circuits of the Fermi–Hubbard, Ising, and Heisenberg models, as the number of magic state distillation factories is increased from 1 to 4.  (\textbf{d}) For the 10x10 Ising circuit, the bar graph shows the effect of varying the magic state processing time.} 
  \label{fig:19}
\end{figure}

 \begin{figure}  
  \centering
  \includegraphics[width=0.5\textwidth]{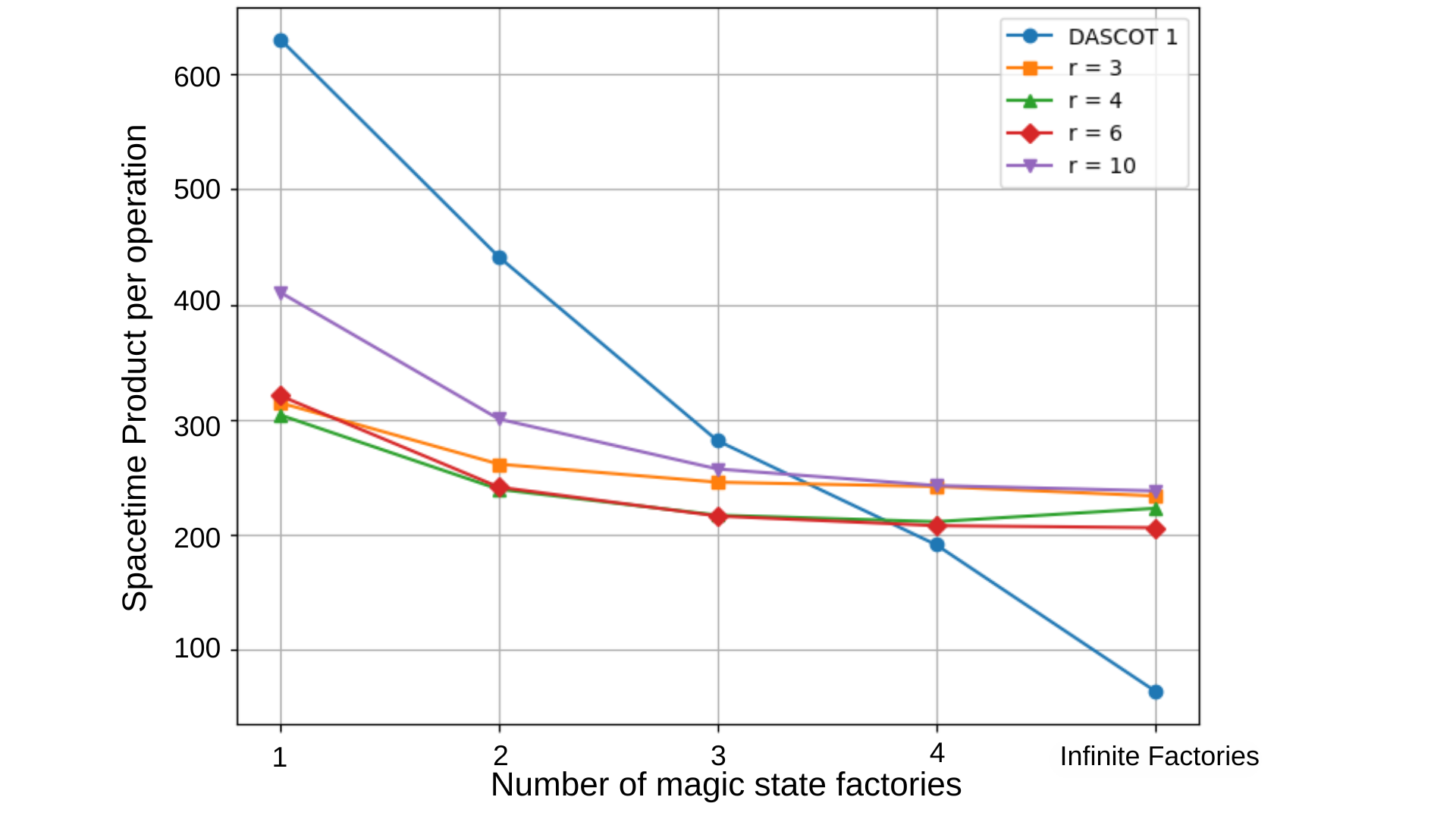}
  \includegraphics[width=0.5\textwidth]{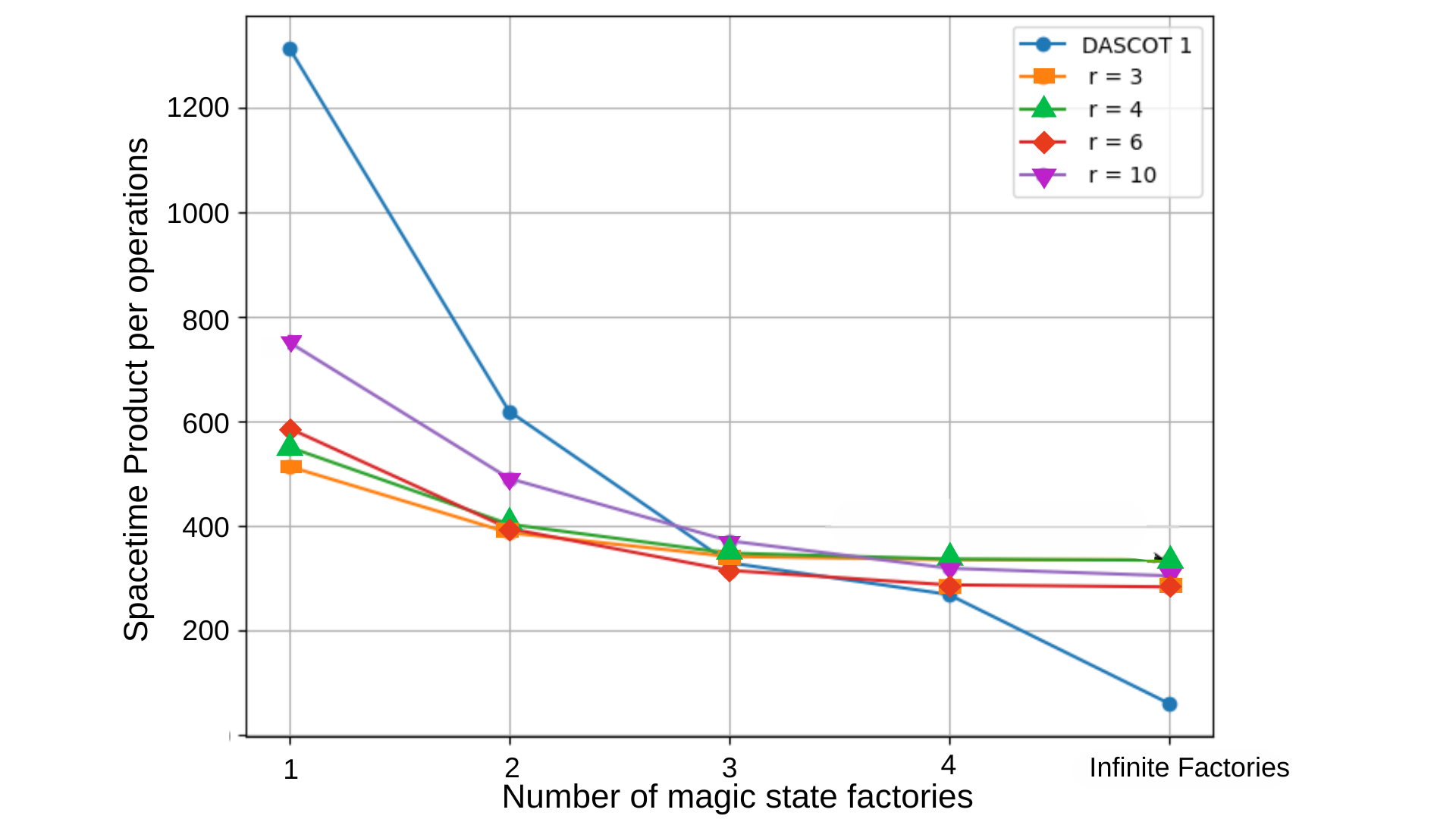}
  \caption{Comparison with DASCOT method of \cite{10.1145/3720416}. Spacetime volume versus the number of distillation factories, with routing paths varied as specified in Figure~\ref{fig:9}. Top: 10x10 Fermi–Hubbard model. Bottom: 10×10 Ising model. The parameter `r' denotes the number of routing paths. In this plot, spacetime volume excludes magic state factories as DASCOT assumes an unlimited supply of magic states.} 
  \label{fig:1117}
\end{figure}

\subsection{Comparison with LSQCA}
In Figure \ref{fig:16}, we compare qubit cost, execution time, and the spacetime volume against the LSQCA approach presented in \cite{10946814}. The referenced work proposes a load/store quantum computer architecture and evaluates two designs, namely the Point SAM (Scan-Access Memory) and the Line SAM. As reported in the paper, the Line SAM architecture reduces latency significantly by increasing the number of scan cells; therefore, we compare our results against this more optimal design. We evaluate the condensed matter quantum circuits and include additional benchmarks from \cite{li2022qasmbench} given in Table \ref{tab:gate_counts}. Except for the GHZ circuit, all other circuits contain T gates and require magic state routing. The execution time is compared using \emph{CPI} (cycles per instruction), defined as the total execution time divided by the number of operations in the input circuit. CPI helps us assess how efficiently the techniques execute instructions independent of the program size. In Figure  \ref{fig:16}, we compare the most optimal layouts for each benchmark for comparisons. The magic state factory is fixed at one, with production time of 11d timesteps.

We further examine the impact of varying the number of factories and the processing time of magic states on the execution time in both approaches. In Figure \ref{fig:19}(a, b, c), we compare the cost per instruction to execute 100 qubit Fermi-Hubbard, Ising and Heisenberg 2D circuits, respectively, with an increase in the number of factories from 1 to 4. The sequential nature of Line SAM prevents a reduction in execution time as the number of factories increases. This is because, while the T state availability is improved with the number of factories, the movement of data qubits between regions takes up a significant amount of time. For the Ising 2D circuit with one factory, the execution time of Line SAM is 1.003$\times$ that of our approach, which increases to 1.69$\times$ when four factories are used. In \ref{fig:19}(d), we study the change in the CPI as we vary the distillation processing times for the Ising circuit. We observe that reducing the processing time leads to a substantial decrease in time overheads for our approach relative to Line SAM. This improvement arises because Line SAM permits considerably less parallelism within the circuit. Consequently, it achieves near-optimal performance only when the magic state processing bottleneck is sufficiently large to dominate all other sequential operations in the architecture.  
 
\subsection{Comparison with DASCOT}
We compare our results with DASCOT \cite{10.1145/3720416}'s results. DASCOT exploits dependency structures between circuit  operations to optimise for mapping and routing problems. This method only deals with routing for two-qubit operations and magic states and generates steps for implementing  these operations from a circuit in parallel. While the approach generates near-optimal results for routing, the layout described in the paper uses 3$\times$ more qubits than our layouts. Figure  \ref{fig:1117} is a plot of the spacetime volume per operation as a function of the number of factories. The two figures correspond to the single Trotter step 10x10  Fermi-Hubbard 2D (top) and Ising 2D (bottom) models while changing the available routing paths. The first entry (blue) is the DASCOT result for the compact layout. DASCOT assumes an unlimited supply of magic states while fixing the number of qubit ports in the layout. It does not incorporate the bottlenecks associated with state distillation when generating the execution steps. Moreover, it does not use ``move'' operations because it is  a compilation technique used to route CNOTs and T states. However, for our comparison,  we introduce the compilation bottleneck as an added constraint in the result and compare for different numbers of factories.  We consider factory counts ranging from 1 to 4; in addition, a fifth data point representing the scenario with infinite factories. Results show that when there is an abundance of T states, the spacetime product of the DASCOT approach is the lowest. At this point, the spacetime product of our approach is, on average, 4.7$\times$ more than that of DASCOT. However, as we introduce the distillation constraints,  DASCOT's spacetime volume increases relative to our results. This is because while DASCOT's routing method is able to parallelise operations through routing, it is unable to perform T gates with a reduced availability of T states due to the architecture constraints. At one factory in the 2D Fermi-Hubbard circuit (top), the DASCOT spacetime product averages at  1.96$\times$ the spacetime product from our approach (averaged over the routing paths).

\section{Conclusion}
Early FTQC systems are expected to be developed in the next five years. Yet, compilation techniques for fault-tolerant quantum computation are focused on the long term with minimum resource constraints. Our work considers the compilation challenge for early FTQC systems and contributes heuristics that offer large improvements in qubit counts and spacetime volumes for important quantum applications compared to prior works. Surprisingly, we show that very simple greedy heuristics come close to optimal performance in these metrics. As systems scale up, our work has the potential to be an important part of the compilation toolchain and inform software design.The code implementing the methods described in this paper is publicly available at \url{https://github.com/sanaasharma/sto-eftqc}.


\balance
\bibliographystyle{IEEEtran}   
\bibliography{References/references}

\end{document}